\documentclass[paper, 12pt, letterpaper, epsf]{JHEP} 
\input{epsf.tex}
\usepackage{graphics} 
\usepackage{epsfig}

\def\Bbb{\bf} 

\def\C{{\Bbb C}} 
\def\R{{\Bbb R}}
\def\Z{{\Bbb Z}}

\def\P{{\Bbb P}}

\def\id{\protect{{1 \kern-.28em {\rm l}}}}

\newcommand{\be}{\begin{equation}} \newcommand{\ee}{\end{equation}}
\newcommand{\bea}{\begin{eqnarray}} \newcommand{\eea}{\end{eqnarray}}
\newcommand{\beann}{\begin{eqnarray*}} \newcommand{\eeann}{\end{eqnarray*}}
\newcommand{\bfig}{\begin{figure}} \newcommand{\efig}{\end{figure}}
\newcommand{\nn}{\nonumber}
\newcommand{\ba}{\begin{array}}\newcommand{\ea}{\end{array}}

\newtheorem{Proposition}{Proposition}[section]

\newtheorem{Theorem}{Theorem}[section]
\newtheorem{Lemma}{Lemma}[section]
\newtheorem{Corrolary}{Corrolary}[section]

\newcommand{\bp}{\begin{Proposition}} \newcommand{\ep}{\end{Proposition}} 
\newcommand{\bt}{\begin{Theorem}} \newcommand{\et}{\end{Theorem}} 
\newcommand{\bl}{\begin{Lemma}} \newcommand{\el}{\end{Lemma}} 
\newcommand{\bc}{\begin{Corrolary}} \newcommand{\ec}{\end{Corrolary}} 

\title{Domain walls of N=2 supergravity in five dimensions 
from hypermultiplet moduli spaces}

\author{L.~Anguelova, C.~I.~Lazaroiu
\\C.~N.~Yang Institute for Theoretical Physics\\
SUNY at Stony Brook, NY11794-3840,
U.S.A.\\anguelov, calin @insti.physics.sunysb.edu}

\abstract{We study domain wall solutions in d=5, N=2 supergravity
coupled to a single hypermultiplet whose moduli space is described by
certain inhomogeneous, toric ESD manifolds constructed recently by
Calderbank and Singer. Upon gauging a generic $U(1)$ isometry of these
spaces, we obtain an infinite family of models whose
``superpotential'' admits an arbitrary number of isolated critical
points. By investigating the associated supersymmetric flows, we prove
the existence of domain walls of Randall-Sundrum type for each member
of our family, and find chains of domain walls interpolating between
various $AdS_{5}$ backgrounds. Our models are described by a 
discrete infinity of smooth and complete one-hypermultiplet moduli
spaces, which live on an open subset of the minimal resolution of
certain cyclic quotient singularities.  These spaces generalize the
Pedersen metrics considered recently by Behrndt and Dall' Agata. }

\preprint{YITP-SB-02-45}

\begin{document}

\tableofcontents

\pagebreak 

\vskip .6in

\section{Introduction}

Five dimensional gauged supergravity has acquired some 
phenomenological interest due to several recent developments. 
The work of \cite{LOSW} showed that the reduction of 
Horava-Witten theory \cite{HW} to five dimensions is a gauged minimal
supergravity admitting a BPS saturated
domain wall solution which can be identified with the four-dimensional 
space-time of a strongly-coupled heterotic compactification \cite{Witten_sh}. 
Another direction is provided by the AdS/CFT correspondence 
\cite{Mald}. In this framework the domain walls of $N=8$ 
gauged supergravity have a natural interpretation as renormalization 
group flows in the corresponding field theory. 
When an embedding of $N=2$ supergravity into the $N=8$ theory is 
known, the associated domain walls of the $N=2$ theory aquire an 
RG flow interpretation\footnote{It has become customary to use 
RG flow terminology even when such an embedding is not known, and we shall 
do so in what follows.}.
Yet another development is the proposal 
of \cite{RS} for an alternative to compactification. This scenario requires a 
domain wall interpolating between two $AdS_5$ solutions (of equal 
vacuum energy density) associated 
with IR points (critical points for the ``superpotential'' 
where the warp factor is exponentially small). 
Despite intense interest in the subject, there 
has been limited progress in finding explicit supergravity realizations
of such scenarios. 

In this regard several no-go theorems were proposed \cite{KL, CvB, GibL,
MN}, which state that, under certain assumptions, there are no
supersymmetric domain wall solutions connecting IR critical points of
the supergravity potential.  As it turns out, the relevant assumptions 
{\em can} be violated once one considers coupling
to hypermultiplets.  In particular, the recent work of \cite{BA}
provides a counterexample obtained by coupling the supergravity
multiplet to a single hypermultiplet described by a certain
non-homogeneous quaternion-Kahler space; in this model, the no-go
theorems of \cite{KL, GibL, MN} do not apply.  This
underscores the importance of reconsidering the problem 
in the general context of inhomogeneous hypermultiplet moduli spaces.

As a general rule, however, one knows quite a bit about flows on the
vector/tensor multiplet moduli space, but rather little about their
hypermultiplet counterpart. The difficulty in the latter case 
consists mainly in understanding the associated geometry.  It is well-known
that the hypermultiplet moduli space must be a quaternion-Kahler space
of negative scalar curvature.  To trust the supergravity
approximation, one must restrict to smooth quaternion-Kahler
spaces\footnote{In principle, one may allow for curvature
singularities in the {\em classical} hypermultiplet moduli space.
However, one expects such singularities to be removed by quantum
effects, for example if the model under consideration can be realized
in string/M-theory.}.
Even restricting to one hypermultiplet (the focus of the present paper), 
very little is known explicitly for the generic case. In this situation, 
the quaternion-Kahler condition is equivalent to the requirement that $M$
is Einstein-selfdual (ESD). The simplest negative curvature examples 
are provided by the homogeneous spaces $SU(2,1)/U(2)$ (the moduli
space of the universal hypermultiplet, i.e. the Bergman metric)
and $EAdS_4=SO(4,1)/SO(4)$ (the Euclidean version of $AdS_4$, also
known as the hyperbolic space $H^4$ or the hyperbolic metric on the
open four-ball).  Another class of examples is provided by
cohomogeneity one $SU(2)$-invariant complete ESD metrics, which were
classified in \cite{Hitchin}. A distinguished subclass of the latter is
provided by those metrics which admit an isometric $U(2)$ action.
These are the Pedersen metrics on the open four-ball \cite{Pedersen}
and their analytic continuations \cite{Pedersen, Hitchin}. As it turns 
out, these are the metrics relevant for 
the counter-example of \cite{BA}\footnote{The authors of \cite{BA} use a
parameterization due to \cite{P}, which is quite different from that
of \cite{CP} and \cite{Pedersen}, and somewhat cumbersome for our
purpose. The relation between their coordinates and those of
\cite{Pedersen} is described in Appendix A.}.

What will allow us to make progress is the recent work of
Calderbank and Pedersen \cite{CP}, which gave an explicit description
of the most general ESD space admitting two commuting and linearly
independent Killing vector fields. Through an elegant chain of
arguments, they showed that such spaces are described by a single
function $F$ of two variables, which is constrained to obey a {\em
linear} PDE (namely, $F$ must be an eigenfunction of the
two-dimensional hyperbolic Laplacian with eigenvalue $3/4$).  An
immediate consequence of this linear description is that one can
obtain new solutions (at least locally) by superposing various
eigenfunctions $F$ --- a situation which is quite unexpected at first
sight.\footnote{Note that an abstract classification of quaternion-Kahler spaces with $n$ quaternionic abelian isometries in terms of a single function was infered in \cite{dWRV}, based on the relation to hyperkahler cones with $n$ abelian triholomorphic isometries. However in \cite{dWRV} the relation between the quaternion-Kahler metric and the single function characterizing it is rather implicit and hence more of conceptual than practical significance.}

For the case of {\em positive} scalar curvature, the work of \cite{CP} has
another application: it leads to an elegant
description of ESD metrics on certain compact toric orbifolds which
include and generalize the models studied a while ago by Galicki and Lawson
\cite{GL}. As explained in \cite{toric} and \cite{metrics}, this can
be combined with the construction of \cite{BS} and \cite{GPP} in order
to produce a large class of conical $G_2$ metrics, which lead to
interesting M-theory backgrounds which produce chiral field theories in four
dimensions \cite{Acharya_Witten}. The main simplification for the
positive curvature case is due to Myers's theorem, which forces such
spaces to be compact (if complete); this makes them amenable to
(hyperkahler) toric geometry techniques upon invoking the associated
hyperkahler cone/Swann bundle. In particular, this allows one to extract
{\em global} information by simple computations in integral linear
algebra.

When studying hypermultiplets, the ESD
space of interest has negative scalar curvature and 
the isometries of $M$ may fail to be
compact.  Therefore, toric geometry techniques do not always apply.
This reflects the well-known observation that the global
study of Einstein manifolds of negative scalar curvature is
considerably more involved than the positive curvature case. In
particular, it is not trivial to find functions $F$ for which the
metric of \cite{CP} is smooth and complete as required by supergravity
applications. A class of such solutions was recently given
by Calderbank and Singer \cite{CS}, and in this paper we shall restrict to
negative curvature models of that type. The metrics of
\cite{CS} are smooth and complete, and live on an open subset $M_+$ of
a toric resolution $M$ of an Abelian quotient singularity $\C^2/\Z_p$;
the set $M_+$ contains all irreducible components of the exceptional
divisor.  To ensure negative scalar curvature, one must require
$c_1(M)<0$\footnote{Compare this with the Gorenstein case $c_1(M)=0$
($A_{p-1}$ surface singularities), which leads to the well-known
hyperkahler metrics of Eguchi--Hanson and Gibbons--Hawking 
\cite{EH}.}. These models
admit two Killing vectors with {\em compact} orbits, and thus they are
invariant under a $U(1)^2$ action.  As pointed out in \cite{CP, CS},
such models are a generalization of the Pedersen--LeBrun metrics
\cite{Pedersen,LeBrun}; 
the latter arise as the particular case when $M={\cal
O}(-p)\rightarrow \P^1$ is the minimal resolution of $\C^2/\Z_p$ with
the {\em symmetric} $\Z_p$ action $(z_1,z_2)\rightarrow (e^{2\pi
i/p}z_1,e^{2\pi i/p} z_2)$.  As mentioned above, the Pedersen metrics
are invariant with respect to a larger $U(2)=SU(2)\times U(1)$ action
and fit into the cohomogeneity one classification of $SU(2)$ invariant
ESD metrics given by Hitchin \cite{Hitchin}.

The main advantage of the metrics of \cite{CS} is that the underlying
manifold admits a toric description, even though the metrics
themselves have negative scalar curvature. Indeed, the resolution $M$
is a (noncompact) toric variety in two complex dimensions, whose
combinatorial description is a classical result \cite{BPV, Oda}. In
particular, the orbits of the isometric $T^2$ action can be described
by standard methods of toric geometry \cite{Danilov, Oda, Fulton,
Audin}.  When considering flows on such spaces, this allows for easy
identification of the critical points of the relevant superpotential
since, in the absence of vector/tensor multiplets, the latter are the fixed
points of the U(1) isometry used to gauge the supergravity action
\cite{CDKP, Cortes}.

In fact, the basic picture can be explained quite easily in
non-technical language.  Recall that the toric variety $M$ can be
presented as a $T^2$- fibration over its Delzant polytope
$\Delta_M$\footnote{This fibration arises by considering the moment
map $\mu:M\rightarrow \R^2$ of the $U(1)^2$ action with respect to the
toric Kahler metric of $M$. The Delzant polytope $\Delta_M$
is the image $\mu(M)\subset \R^2$. Note that the toric Kahler
metric of $M$ differs from its ESD metric.}. In our
case, the latter is a noncompact planar polytope and general results
\cite{Oda, Audin} show that the $T^2$ fiber of $M$ collapses to a
point at its vertices and to a circle above its edges. The 
$S^1$ fibrations above the finite edges (whose circle fibers collapse to
points at the vertices) give a collection of smooth spheres $S_j$
which are holomorphically embedded in $M$ --- one obtains a copy of
$\P^1$ for every finite edge of $\Delta_M$. A generic isometry fixes only the
points $p_j$ of $M$ sitting above the vertices of $\Delta_M$. As one
can obtain an arbitrary number of points $p_j$ by taking $p$ ($M$ is
the resolution of $\C^2/\Z_p$) to be large, {\em one can produce
models with an arbitrarily large number of isolated critical points of
the superpotential}. This should be contrasted with the Pedersen
metrics considered in \cite{BA}, which lead to at most two isolated
critical points. This observation will allow us to build chains of
flows connecting the critical points, and therefore domain wall
solutions which interpolate between the associated $AdS_5$
backgrounds.

For a general choice of gauged isometry, it turns out that at most one
such flow is of Randall-Sundrum type (i.e it connects two IR critical
points).  Among the rest, there are domain wall solutions which
interpolate between a UV and an IR critical point. Some of these may
describe RG flows of appropriate dual field theories due to the
following chain of arguments. It is believed that 5-dimensional $N=8$
gauged supergravity \cite{GRW, PPN} is a consistent truncation of the
10-dimensional IIB supergravity on $AdS_5\times S^5$ (some evidence
for this was presented in \cite{KPW, Ca}). This means that every
solution of the former is also a solution of the latter. Although
there is no proof at present, this is strongly supported by similarity
with two other cases: 4-dimensional $N=8$ gauged supergravity, which
is known \cite{WN} to be a consistent truncation of 11-dimensional
supergravity on $AdS_4\times S^7$, and 7-dimensional gauged
supergravity, which was shown recently \cite{NVN} to be a consistent
truncation of 11-dimensional supergravity on $AdS_7\times
S^4$. Additional, indirect evidence for the consistency of the
truncation of IIB SUGRA on $AdS_5\times S^5$ to $d=5$, $N=8$ SUGRA is
provided by the numerous studies in $AdS/CFT$ (for example \cite{FGPW,
GPPZ}), where that consistency is assumed and domain walls of the
supergravity theory are interpreted as RG flows of the corresponding
dual field theory. Various quantities, calculated both from the
gravity side and from the field theory side, have been successfully
matched \cite{FGPW, GPPZ}. For one of these solutions --- a domain
wall in 5d $N=8$ supergravity, which describes an RG flow from $N=4$
to $N=1$ super Yang-Mills driven by the addition of a mass term for
one of the three adjoint chiral superfields \cite{FGPW} --- it was
found in \cite{CDKP} that it can be embedded in 5d $N=2$ gauged
supergravity. In particular this means that the $N=2$ theory is at
least in that case a consistent truncation of the 10d IIB theory. This
may be by chance, but it may also be that many more domain walls of 5d
$N=2$ SUGRA can be embedded in the 10d theory and hence can have via
the $AdS/CFT$ correspondence an interpretation as a RG flow of an
appropriate field theory. It would be interesting to explore whether
some of our UV-IR flows have such interpretations.

The present paper is organized as follows. In Section 2 we recall the
necessary ingredients of 5-dimensional $N=2$ gauged supergravity, 
and extract the superpotential and flow
equations relevant for coupling the gravity multiplet to
a single hypermultiplet described by the metric of
\cite{CP}. Section 3 describes the subclass of metrics constructed in 
\cite{CS}. In Section 4 we study
the general properties of BPS flows for such models. 
In particular, we give the general flow solutions connecting our 
critical points and describe the conditions under which such a solution
has Randall-Sundrum type (i.e. is an IR-IR flow). Section 5
illustrates this discussion with a few explicit examples, 
which include and generalize the solution of \cite{BA}. Appendix 
A gives the coordinate transformation relating the Pedersen metrics 
to the models discussed in \cite{BA}.

\section{Flow equations on toric one-hypermultiplet moduli spaces} 

Consider coupling a single hypermultiplet to the supergravity
multiplet in five dimensions.  As the theory contains only one gauge
field (namely the graviphoton), one can gauge at most one isometry of
the hypermultiplet moduli space.  The general Lagrangian of gauged
minimal supergravity in five dimensions was derived in
\cite{CD}\footnote{Minimal means $N=2$ as in 5d there is no $N=1$
gauged supergravity theory.}.  When no vector/tensor multiplets are
present\footnote{When including vector/tensor multiplets, the potential can
not always be written in this form. See \cite{CDKP} for details.}, the
scalar potential induced by the gauging takes the form \cite{CDKP}:
\be 
\label{pot} {\cal V} = - 6 W^2 + \frac{9}{2} g^{XY} 
\partial_X W \partial_Y W \, , 
\ee 
where $g_{X Y}$ is the hypermultiplet metric and the "superpotential" $W$ 
is given by: 
\be
\label{sp} W = \sqrt{\frac{2}{3} P^r P^r} \qquad , \qquad r=1,2,3 \, .
\ee 
Here $\{P^r\}_{r=1}^3$ is the triplet of prepotentials related to the
Killing vector $K^Y$ of the gauged isometry:
\be
\label{KPrel} {\cal R}_{X Y}^r K^Y = 2 D_X P^r \, , 
\ee 
where $\{{\cal R}^r\}_{r=1}^3$ is the triplet of Sp(1) curvatures and
$D_X$ is the full covariant derivative (including the Levi-Civita,
Sp($n$) and Sp(1) connections for the case of a $4n$-dimensional
quaternionic space). The factor of $2$ is a result of a different
normalization w.r.t. \cite{CDKP, BA} as will be explained below.

Note that we work with the convention \cite{CDKP} that $W$ is always
non-negative (i.e. we choose the non-negative square root in
(\ref{sp})). This implies that $W$ will fail to be differentiable at a
zero of $W$ where the directional derivatives do not all vanish.  At
such a {\em noncritical zero}, the function $W^2$ is differentiable,
and its Hessian is positive semidefinite \cite{Cortes}.

In our example, we take $n=1$ and let the hypermultiplet moduli space
be described by a $T^2$ invariant ESD metric of negative scalar
curvature.  As shown in \cite{CP}, the most general $T^2$ invariant
ESD metric has the form:
{\footnotesize \bea \label{metric}
\!\!\!\!\!\!\!\!\!\!  d\sigma^2=\frac{ |F^2 - 4\rho^2(F_\rho^2 +
F_\eta^2)| }{4 F^2}\; \frac{d\rho^2 + d\eta^2}{\rho^2} +\frac{ \left[
(F - 2 \rho F_\rho) \alpha - 2 \rho F_\eta \beta \right]^2 + \left[
-2\rho F_\eta \alpha + (F + 2 \rho F_\rho )\beta \right]^2 }{
F^2|F^2-4\rho^2(F_\rho^2 + F_\eta^2)|}
\eea} 
where $\alpha=\sqrt\rho\,d\phi$, $\beta=(d\psi+\eta\,
d\phi)/\sqrt\rho$ and the function $F(\rho,\eta)$ satisfies the equation: 
\be 
\label{lap}
\rho^2(F_{\rho \rho} + F_{\eta \eta}) = \frac{3}{4} F \, .  
\ee 
Note that we use the notation $F_\rho= \partial_\rho F,
F_\eta=\partial_\eta F$, $F_{\rho\rho}= \partial_\rho^2 F$ etc. for
the partial derivatives of $F$.

In equation (\ref{metric}), one takes $\rho>0$ and $\eta\in \R$, while
$\phi,\psi$ are coordinates of periodicity $2\pi$.  The metric is
well-defined for $F^2\neq 4\rho^2(F_\rho^2+F_\eta^2)$ and $F\neq 0$.
It has positive scalar curvature in the regions where $F^2 -
4\rho^2(F_\rho^2+F_\eta^2)>0$ and negative scalar curvature for $F^2 -
4\rho^2(F_\rho^2+F_\eta^2)<0$. One can easily check that
(\ref{metric}) is normalized so that in the latter case the scalar
curvature is $-12$ as is usual for supergravity applications.

Condition (\ref{lap}) says that $F$ is an eigenfunction (with
eigenvalue $3/4$) of the hyperbolic Laplacian $\Delta_{\cal
H}=\rho^2(\partial_\rho^2+\partial_\eta^2)$.  This is the Laplacian of
the standard metric:
\be
ds^2_{\cal H}=\frac{1}{\rho^2}(d\rho^2+d\eta^2)
\ee
on the hyperbolic plane
${\cal H}^2$ with coordinates $\rho>0$ and 
$\eta\in \R$. Note that we use the upper 
half plane model.

The Sp(1) curvatures determined by (\ref{metric}) are \cite{CP}: 
\bea
\label{curv} {\cal R}^1 &=& - \, \frac{F^2 -4 \rho^2 (F_{\rho}^2 +
F_{\eta}^2)}{4 F^2 \rho^2} \, d\rho \wedge d\eta + \frac{1}{F^2} \,
d\phi \wedge d\psi \nn \\ {\cal R}^2 &=& \frac{F_{\eta}}{F^2
\sqrt{\rho}} \, d\psi \wedge d\rho + \frac{1}{F^2 \sqrt{\rho}}
\left(\rho F_{\rho} + \eta F_{\eta} - \frac{F}{2} \right) d\phi \wedge
d\rho \nn \\ &-& \frac{1}{F^2 \sqrt{\rho}} \left( F_{\rho} +
\frac{F}{2 \rho} \right) \, d\psi \wedge d\eta + \frac{1}{F^2
\sqrt{\rho}} \left( \rho F_{\eta} - \eta F_{\rho} - \frac{\eta}{\rho}
\frac{F}{2} \right) \, d\phi \wedge d\eta \nn \\ {\cal R}^3 &=& - \,
\frac{1}{F^2 \sqrt{\rho}} \left(F_{\rho} + \frac{F}{2 \rho}\right)
d\psi \wedge d\rho + \frac{1}{F^2 \sqrt{\rho}} \left( \rho F_{\eta} -
\eta F_{\rho} - \frac{\eta}{\rho} \frac{F}{2} \right) d\phi \wedge
d\rho \nn \\ &-& \frac{F_{\eta}}{F^2 \sqrt{\rho}} \, d\psi \wedge
d\eta - \frac{1}{F^2 \sqrt{\rho}} \left( \rho F_{\rho} + \eta F_{\eta}
- \frac{F}{2} \right) d\phi \wedge d\eta \, .  
\eea
 
It is easy to check that they satisfy: 
\be 
\label{RR} {\cal R}_{XY}^r {\cal R}^{sYZ}
= - \delta^{rs} \delta_X^Z - \varepsilon^{rst} {\cal R}_X^t{}^Z 
\ee
unlike (2.11) of \cite{CDKP}. Hence these curvatures are normalized to
\be 
{\cal R}_{XY}^r = - J_{XY}^r \, , 
\ee 
where $J^r$ are the three complex structures, and not to 
${\cal R}^r = - \frac{1}{2} J^r$ as in
\cite{CDKP, BA}. With this normalization the covariant derivative
takes the form 
\be
\label{Sp1CovDer} D_X P^r = \partial_X P^r + \varepsilon^{rst}
\omega_X^s P^t   
\ee
on a quantity having only an Sp(1) index.  Note the slight difference
w.r.t. (2.12) of \cite{CDKP}. In (\ref{Sp1CovDer})
$\{\omega^s\}_{s=1}^3$ is the triplet of Sp(1) connections for the
curvature (\ref{curv}) \cite{CP}:
\be 
\label{con} \omega^1 = - \frac{F_{\eta}}{F} d\rho +
\left(\frac{1}{2 \rho} + \frac{F_{\rho}}{F} \right) d\eta~~,~~\omega^2
= - \frac{\sqrt{\rho}}{F} d\phi~~,~~\omega^3 =
\frac{\eta}{F\sqrt{\rho}} d\phi + \frac{1}{F \sqrt{\rho}} d\psi \, .
\ee 
Using (\ref{RR}) one can solve (\ref{KPrel}) for the Killing
vector: 
\be 
\label{Kil} K^Y = - \frac{2}{3} {\cal R}^{rYX} D_X P^r \, .  
\ee 
Again the numerical factor differs slightly from \cite{CDKP,BA} due to
the different normalization of the Sp(1) curvatures and
connections. Equation (\ref{KPrel}) can also be solved for the
prepotentials by using the fact that they are eigenfunctions of the
Laplacian \cite{DF}:
\be 
D^X D_X P^{\, r} = - \, 2n P^{\, r} 
\ee 
(for a $4n$-dimensional quaternion-Kahler space).  Again this differs
by a factor of $2$ w.r.t. the footnote after (2.15) in
\cite{CDKP}. One obtains: 
\be
\label{prepot} P^{\, r} = - \frac{1}{4} D_X (K^Z g_{ZY} {\cal R}^{r
XY}) = - \frac{1}{4} {\cal R}^{r X Y} \partial_X K_Y \, , 
\ee 
where the second equality results from the covariant constancy and
antisymmetry in $X$ and $Y$ of the Sp(1) curvature. A domain wall
of gauged supergravity, 
\be 
\label{ansatz}
ds^2 = e^{2 U(t)} dx_{4d}^2 +
dt^2 \, , 
\ee 
which preserves $N=1$ supersymmetry is given by the solution of the
following system \cite{CDKP}:
\bea
\label{flow} 
\partial_t U &=& \pm g W \nn \\ \partial_t q^X &=& \mp 3 
g g^{XY} \partial_Y W \, , 
\eea 
where $q^X$ are the hypermultiplet scalars. The signs in these
equations must be chosen consistently (i.e. one must use the minus
sign in the second equation if one chooses the plus sign in the
first).  In order to insure continuity of the derivative of a flow
which passes through a noncritical zero of $W$, one must switch the
sign in the first equation when the flow meets such a point.
Accordingly, the sign in the second equation must also be switched
there.

Let us gauge the isometry of the metric (\ref{metric}) along the
Killing vector:
\be 
\label{K} 
K = \partial_{\phi} -\lambda \,\partial_{\psi} \, .  
\ee 
Note that we normalize $K$ such that the coefficient of its
$\partial_\phi$ component equals one.

\

\paragraph{Observation}

Equation (\ref{KPrel}) fixes the prepotentials $P^r$ (and thus the
superpotential (\ref{sp})) in terms of a specific choice for the
Killing vector $K$. In particular, rescaling $K$ leads to a rescaling
of $W$, which can be absorbed by a rescaling of $t$ in equations
(\ref{flow}). Since the first of these equations determines $U$ only
up to a constant factor, this rescaling of $t$ can be further absorbed
into a constant rescaling of the metric (\ref{ansatz}), upon choosing
an appropriate integration constant for $U$.  In particular, the
normalization in (\ref{K}) amounts to a particular choice of scale for
$t$ or, equivalently, a choice of scale for the metric (\ref{ansatz}).

\

Equation (\ref{prepot}) implies: 
\be 
P^1 = 0 \, , \qquad P^2 = \frac{1}{2} \frac{
\sqrt{\rho}}{F} \, , \qquad P^3 = - \frac{1}{2} \frac{\eta -\lambda}{F
\sqrt{\rho}} \,   
\ee
where we took (\ref{lap}) into account.  Using (\ref{Kil}), one can
check that these prepotentials give (\ref{K}). Now (\ref{sp}) gives:
\be 
\label{W}
W = \sqrt{\frac{1}{6 F^2} \left[ \rho + \frac{(\eta -
\lambda)^2}{\rho} \right]} \, .  
\ee 

Because $W$ is only a function of $\rho$ and $\eta$ and the metric
(\ref{metric}) (and hence its inverse) is diagonal in $\rho, \eta$,
the potential (\ref{pot}) and the "flow equations" (\ref{flow})
acquire a particularly simple form:

\be 
{\cal V} = - \frac{1}{F^2} \left( \rho + \frac{(\eta -
\lambda)^2}{\rho} \right) + \frac{18 \rho^2 F^2}{|F^2 - 4
\rho^2(F_{\rho}^2 + F_{\eta}^2)|} [(\partial_{\rho} W)^2 +
(\partial_{\eta} W)^2] \, ,
\ee
and: 
\bea 
\label{Flow}
\frac{d \phi}{d t} &=&\frac{d \psi}{d t} =0\nn\\ \frac{d U}{d t} &=&
\pm g W \nn \\ \frac{d \rho}{d t} &=& \mp 12 g \frac{\rho^2 F^2}{|F^2
- 4 \rho^2(F_{\rho}^2 + F_{\eta}^2)|} \, \partial_{\rho} W \nn \\
\frac{d \eta}{d t} &=& \mp 12 g \frac{\rho^2 F^2}{|F^2 - 4
\rho^2(F_{\rho}^2 + F_{\eta}^2)|} \, \partial_{\eta} W \,. 
\eea

Hence the $T^2$ isometry of the one-hypermultiplet moduli space allows
us to reduce the four-dimensional flow equations to a two-dimensional
problem.  The last two relations describe the gradient flow of $W$
with respect to the metric $\frac{|F^2 - 4 \rho^2(F_{\rho}^2 +
F_{\eta}^2)|}{\rho^2 F^2}(d\rho^2+d\eta^2)$ on the upper half plane
(notice that this is conformal to the hyperbolic metric).

\section{Calderbank-Singer spaces}

\subsection{Minimal resolutions of cyclic quotient singularities}

Consider a cyclic singularity $\C^2/\Z_p$, where the generator of
$\Z_p$ acts through:
\be
(z_1,z_2)\rightarrow (e^{2\pi i/p}z_1, e^{2\pi i q/p} z_2)~~.
\ee
We assume that the integers $p,q$ satisfy $p>q>0$\footnote{ This
action embeds diagonally in $U(2)$.  It embeds in $SU(2)$ if and only
if $q+1=p$, when the singularity $\C^2/\Z_p$ is called {\em
Gorenstein} and has trivial dualizing sheaf; in that case, it is
simply an $A_{p-1}$ surface singularity. In this paper, we are
emphatically {\em not} interested in the Gorenstein case. We note that
non-Gorenstein cyclic singularities arise naturally in the study of
normal complex surfaces--this generalizes the better known case of
$A_{p-1}$ singularities, which give local descriptions for the
singularities of $K3$. The minimal resolution of an $A_{p-1}$
singularity has trivial first Chern class and carries the
multi-Eguchi-Hanson metric, which is hyperkahler. As shown in
\cite{CS}, such resolutions {\em never} carry a toric SDE metric of
negative scalar curvature.}. We consider the {\em minimal} resolution
of this singularity, which is a smooth algebraic surface $M$
birational with $\C^2/\Z_p$ and containing no $-1$ curves. If
$S_1\dots S_k$ denote the irreducible components of its exceptional
divisor, then it is a classical fact \cite{BPV} that the intersection
matrix of these components has the form:
\be
\label{imatrix}
(S_i\cdot S_j)=\left[\begin{array}{ccccc}-e_1&1&0&\cdots
&0\\1&-e_2&1&\cdots &0 \\0&1&-e_3&\cdots
&0\\\cdots&\dots&\cdots&\cdots&\cdots\\0&0&0&\cdots&-e_k\end{array}
\right]~~,
\ee
where the diagonal entries are integers satisfying $e_j\geq 2$. The
adjunction formula shows that $c_1(M)\leq 0$, with $c_1(M)<0$ if and
only if all $e_j\geq 3$ and $c_1(M)=0$ iff all $e_j=2$; the latter
case corresponds to $q=p-1$ (the $A_{p-1}$ Gorenstein
singularity). For what follows, we shall consider exclusively the case
$c_1(M)<0$.

It is well known that both $\C^2/\Z_p$ and its minimal resolution are
toric varieties (see, for example, \cite{Oda}). As explained in
\cite{Oda}, the toric description of $X$ can be extracted with the
help of continued fractions.  Indeed the integers $k$ and $e_1\dots
e_k$ are given by the {\em minus}\footnote{This differs from the more
common `plus' continued fractions. By definition of the expansion
(\ref{cf}), the integers $e_j$ are required to satisfy $e_j\geq 2$.}
continued fraction expansion:
\be
\label{cf}
\frac{p}{q}=e_1-\frac{1}{e_2-\frac{1}{e_3-\dots 
\frac{1}{e_{k-1}-\frac{1}{e_k}}}}~~,
\ee
which we shall denote by $(e_1\dots e_k)$ for simplicity. The toric
data of $M$ can be determined as follows \cite{Oda}\footnote{Our
presentation differs from that of \cite{Oda} in a few trivial
ways. First, reference \cite{Oda} uses a different description of the
cyclic action, which amounts to the redefinitions $p\rightarrow q$ and
$q\rightarrow q-p$.  It also writes our second order recursion as a
first order recursion for two vectors.}.  Consider a basis $(t,t')$ of
the two-dimensional lattice $\Z^2$, and define vectors $\nu_0\dots
\nu_k$ by the two-step recursion:
\be
\label{vec_recursion}
\nu_{j+1}=e_j \nu_j-\nu_{j-1}~~(j=1\dots k)~~,
\ee
with the initial conditions $\nu_0=t$ and $\nu_1=t+t'$. Then
$\nu_{k+1}=(p-q)t+pt'$ and $\nu_0\dots \nu_{k+1}$ are the toric
generators of the minimal resolution $M$, while $\nu_0$ and
$\nu_{k+1}$ are the toric generators of the singularity
$\C^2/\Z_p$. The latter generate a (strongly convex) cone $\sigma$,
subdivided by the vectors $\nu_1\dots \nu_k$ which lie in its
interior. In fact, these vectors coincide with the vertices of the
convex polytope defined by the convex hull of the intersection of
$\sigma-\{0\}$ with the $\Z^2$ lattice.  Following \cite{CS}, we
choose $t=\left[\begin{array}{c}0\\-1\end{array}\right]$ and
$t'=\left[\begin{array}{c}1\\1\end{array}\right]$ (always possible via
a modular transformation), which gives
$\nu_0=\left[\begin{array}{c}0\\-1\end{array}\right]$,
$\nu_1=\left[\begin{array}{c}1\\0\end{array}\right]$ and
$\nu_{k+1}=\left[\begin{array}{c}p\\q\end{array}\right]$.  Upon
writing $\nu_j=\left[\begin{array}{c}m_j\\n_j\end{array}\right]$,
relation (\ref{vec_recursion}) becomes:
\be
\label{mn_recursion}
m_{j+1}=e_j m_j-m_{j-1}~~, \qquad
n_{j+1}=e_j n_j-n_{j-1} \qquad (j\geq 1)~~,
\ee
with the initial conditions $(m_0,n_0)=(0,-1)$ and
$(m_1,n_1)=(1,0)$. These can be recognized as the standard recursion
relations for the numerator and denominator of the partial quotients
$q_j=(e_1\dots e_j)=m_{j+1}/n_{j+1}$ ($j=1\dots k$) of the continued
fraction (\ref{cf}). We remind the reader that the solutions of this
recursion have the following properties (all of which can be checked
by direct computation or induction):

(a) $n_0=-1<n_1=0<n_2=1<n_3<\dots <n_{k+1}=q$ and
$m_0=0<m_1=1<m_2=e_1<\dots <m_{k+1}=p$

(b) $q_1> q_2>\dots > q_k=p/q$~~.

(c) $m_j n_{j+1}-m_{j+1}n_j=1$ for $j=0\dots k$ and
$m_{j-1}n_{j+1}-m_{j+1}n_{j-1}=e_j$ for $j=1\dots k$.

(d) If all $e_j$ are strictly greater that two, then
$m_{j+1}-m_j>m_j-m_{j-1}$ for all $j=1\dots k$ and
$n_{j+1}-n_j>n_{j}-n_{j-1}$ for all $j=2\dots k$.

The first part of (c) says that the area of the triangle determined by
vectors $\nu_j$ and $\nu_{j+1}$ equals $1/2$ --- this is the condition
that the subdivision of the cone $\sigma$ resolves the singularities
of $\C^2/\Z_p$.

\begin{figure}[hbtp]
\begin{center}
\scalebox{0.5}{\input{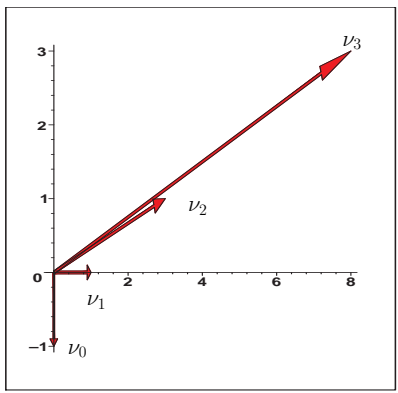}}
\end{center}
\caption{\label{83gens} Toric generators for the model $(p,q)=(8,3)$. In this 
case, one has $k=2$.}
\end{figure}

The situation for the vectors $\nu_0\dots \nu_{k+1}$ is illustrated in
figure \ref{83gens}.  It shows the case $p=8$ and $q=3$, which will be
discussed in more detail in Subsection 5.3.

\subsection{The Calderbank-Singer metrics}

In \cite{CS}, Calderbank and Singer construct toric ESD metrics of
negative scalar curvature on the minimal\footnote{In fact, the
construction of \cite{CS} applies to a more general class of toric
resolutions of $\C^2/\Z_p$, which are not necessarily minimal. In this
paper, we shall consider their construction only for the case of
minimal resolutions.}  resolutions of cyclic singularities with
negative first Chern class.  These metrics are invariant with respect
to the natural $T^2$ action on $M$ induced by its structure of toric
variety. In view of the results of \cite{CP}, they must have the
general form (\ref{metric}) with $F^2 < 4 \rho^2 (F_{\rho}^2 +
F_{\eta}^2)$, where the angular coordinates $(\phi,\psi)$ parameterize
the $T^2$ fibers of $M$.  This is achieved by choosing $F$ to be a
superposition of {\em elementary eigenfunctions} of the type:
\be
f(\rho, \eta; y)=\sqrt{\rho+\frac{(\eta-y)^2}{\rho}}~~,
\ee
which are easily seen to satisfy $\Delta_{{\cal H}}f=\frac{3}{4}f$. 
More precisely, one must choose the linear combination:
\be
\label{Fcs}
F(\rho,\eta)=\sum_{j=0}^{k+1}{w_jf(\rho,\eta; y_j)}~~,
\ee
where: 
\be
\label{yw_def}
w_j=\frac{1}{2}(m_j-m_{j+1})~~,~~y_j=\frac{n_{j+1}-n_j}{m_{j+1}-m_j}
\ee
and we defined $m_{k+2}=0$ and $n_{k+2}=1$ (addition of
$\nu_{k+2}:=\left[\ba{c} m_{k+2}\\n_{k+2}\ea\right]$ and thus of $y_{k+1}$ 
amounts to taking
the one-point compactification ${\overline M}$ of $M$). Since we
assume $c_1(M)<0$, we have $e_j\geq 3$ for all $j=1\dots k$ and thus
$w_{k+1}=p/2>0>w_0=-1/2>w_1>\dots >w_k$.  The combination (\ref{Fcs})
is fixed (up to a constant scale factor) by the requirement that the
metric (\ref{metric}) extends smoothly to the singular fibers of
$M$. When considering the metric (\ref{metric}), one identifies
topologically the boundary of the Delzant polytope $\Delta_M$ with the
boundary $\rho=0$ of the upper half plane model of the hyperbolic
plane. Then the vertices of the Delzant polytope are mapped to the
points $y_0\dots y_{k}$, and its edges correspond to the intervals
determined by these vertices. It is easy to check \cite{CS} that
$y_0>y_1>\dots >y_k>q/p>y_{k+1}$, so that the edges of $\Delta_M$
correspond to the intervals $I_j:=(y_j, y_{j-1})$ sitting at $\rho=0$.

\begin{center}
\vbox{
\begin{minipage}[t]{3in}
\begin{center}
\scalebox{0.4}{\input{plane.pstex_t}}
\end{center}
\end{minipage}~~~~~\begin{minipage}[t]{3in}
\begin{center}
\scalebox{0.4}{\input{M.pstex_t}}
\end{center}
\end{minipage}

\begin{minipage}[t]{3in}
\begin{center}
{\bf Figure 2:} {\footnotesize General arrangement of $y_j$, $Z$ and
$\eta_0=q/p$ in the hyperbolic plane. The simple curve $Z$ separates
the upper half plane into the regions $D_-$ and $D_+$. The latter is
the region of interest for the present paper.}
\end{center}
\end{minipage}~~~~~\begin{minipage}[t]{3in} \begin{center}
{\bf Figure 3:} {\footnotesize The open sets $M_+$ and $M_-$ 
are separated by the conformal infinity $Y$.
Each region $M_{\pm}$ carries an ESD metric of negative scalar
curvature.  The point $p_{k+1}$ (which lies above $y_{k+1}$) is an
orbifold point; it is the `point at infinity' of ${\overline M}$. 
The figure also
shows the irreducible components of the exceptional divisor, which
connect the points $p_0\dots p_k$ lying above $y_0\dots y_k$.}
\end{center}
\end{minipage}}
\end{center}

Expression (\ref{metric}) determines an ESD metric (of negative scalar
curvature) on the space ${\overline M}\rightarrow 
{\overline {\cal H}^2}$, where
${\overline {\cal H}}^2$ is the conformal  compactification of the
hyperbolic plane (obtained by adding the point at 
infinity).  However, this metric is ill-defined along the set
$Z$ given by the equation $F(\rho,\eta)=0$.  As explained in
\cite{CS}, this locus is a smooth simple curve which intersects the
boundary of ${\overline {\cal H}}^2$ in the points $\eta_0=q/p=1/q_k\in
(y_{k+1},y_k)$ and $\infty$ (see figure 2).  This curve separates
${\overline {\cal H}}^2$ into connected components $D_+$ (defined by
the condition $F(\rho,\eta)>0$) and $D_-$ (defined by
$F(\rho,\eta)<0)$, which pull back to two disjoint open subsets $M_+$
and $M_-$ of ${\overline M}$, 
separated by a region $Y$ defined as the pull-back
of $Z$. The piece $D_+$ contains the points $y_0\dots y_k$, while
$D_-$ contains the point $y_{k+1}$.  The set $Y$ is a compact $T^2$
fibration over $Z$, which coincides topologically with the Lens space
$S^3/\Z_p$. It is a conformal
infinity for each of the two ESD metrics determined by (\ref{metric})
on the open sets $M_+$ and $M_-$ \cite{CP}.  Note that $M_+$ contains
the exceptional divisor of the resolution, and that the ESD metric
induced on $M_+$ is smooth and complete.  The metric induced on
$M_-$ is a complete orbifold metric, with the orbifold point 
given by the point at infinity of ${\overline M}$, which we 
denote by $p_{k+1}$ (this points sits above $y_{k+1}$). 
Since we are interested in smooth and complete metrics,
we shall concentrate on the region $M_+$ (see figure 3).

\subsection{Fixed points of $U(1)$ isometries}

The fibration of $M$ over its Delzant polytope translates into a
fibration over  ${\cal H}^2$. Since
the points $y_j$ correspond to the vertices of $\Delta_M$, the $T^2$
fibers collapse to points above $y_j$ and to circles above the
intervals $I_j=(y_j,y_{j-1})$ sitting at $\rho=0$. 
In expression (\ref{metric}), one uses
coordinates $(\phi,\psi)$ along the $T^2$ fibers such that the
vectors $\partial_\psi,\partial_\phi$ correspond to the canonical
basis of the lattice $\Z^2$. Hence the $U(1)$ generator (\ref{K})
corresponds to the two-vector
$\tau=\left[\ba{c}-\lambda\\\ 1\ea\right]$. This isometry fixes the
sphere $S_j$ lying above $I_j$ precisely when $\tau$ is orthogonal
to the generator $\nu_j$, i.e. when
$\lambda=\lambda_j:=\nu_j^2/\nu_j^1=n_j/m_j=1/q_{j-1}$.  Note that all
isometries (\ref{K}) fix the points $p_j$ of $M$ lying above $y_j$.

\section{Supersymmetric flows on Calderbank-Singer spaces}

\subsection{Critical points of the superpotential}

When the supergravity multiplet is coupled only to hypermultiplets but
not to vector/tensor multiplets, it was shown in \cite{CDKP} and
\cite{Cortes} that the critical points of the superpotential (\ref{W})
coincide with the fixed points of the associated isometry
(\ref{K}). In view of the discussion above, we find that an isometry
of $M_+$ with $\lambda$ different from $\lambda_1\dots \lambda_k$
fixes exactly the points $p_0\dots p_k$; thus a generic isometry has
$k+1$ critical points. In the non-generic cases $\lambda=\lambda_j$,
the isometry fixes $p_0\dots p_k$ together with the entire sphere
$S_j$.

\subsection{Asymptotic form of the flow equations and divisorial flows}
The superpotential (\ref{W}) can be written:
\be
W=\frac{1}{\sqrt 6}\frac{f(\rho,\eta,\lambda)}{F(\rho,\eta)}~~,
\ee
where we used $F>0$ on the domain of interest $D_+$.
Let us write the flow equations (\ref{Flow}) as:
\bea
\label{FLOW}
\frac{dU}{dt}&=&\pm g W\nn\\
\frac{d\rho}{dt}&=&\pm h(\rho,\eta)\nn\\
\frac{d\eta}{dt}&=&\mp g(\rho,\eta)~~,
\eea
where: 
\bea
h(\rho,\eta)&=&
12 g \frac{\rho^2 F^2}{F^2 - 4 \rho^2(F_{\rho}^2 + F_{\eta}^2)} \,
\partial_{\rho} W \nn \\ 
g(\rho,\eta)&=&
- 12 g \frac{\rho^2 F^2}{F^2 - 4 \rho^2(F_{\rho}^2 + F_{\eta}^2)} \,
\partial_{\eta} W~~. 
\eea
It is not very hard to check the following asymptotics for $h,g$ as 
$\rho\rightarrow 0$:
\bea
h(\rho,\eta)&=& g\sqrt{\frac{3}{2}}\frac{\Phi-(\eta-\lambda)^2\Theta}
{(\Xi^2-\Phi\Theta)|\eta-\lambda|}\rho +O(\rho^2)\nn\\
g(\rho,\eta)&=& g\sqrt{\frac{3}{2}}\frac{(\Phi-(\eta-\lambda)\Xi)
sign(\eta-\lambda)}{\Xi^2-\Phi\Theta}+O(\rho)~~.
\eea
To arrive at these expressions, we defined:
\bea
\label{PKT}
\Phi(\eta)&=&\sum_{j=0}^{k+1}{w_j|\eta-y_j|}\nn\\
\Xi(\eta)&=&\sum_{j=0}^{k+1}{w_jsign(\eta-y_j)} \nn\\
\Theta(\eta)&=&\sum_{j=0}^{k+1}{\frac{w_j}{|\eta-y_j|}}~~.
\eea
In particular, one has $\lim_{\rho\rightarrow0^+}{h(\rho,\eta)}=0$, so
that the gradient lines of $W$ become orthogonal to the real axis for
$\rho\rightarrow 0$. Thus one can find a flow (integral curve) along
this axis by setting $\rho=0$ consistently in equations
(\ref{FLOW}). In this case, the second equation in (\ref{FLOW}) is
trivially satisfied and the system reduces to:
\bea
\label{bd_flow0}
\frac{dU}{dt}&=&\pm g W(0,\eta)\nn\\
\frac{d\eta}{dt}&=&\mp g_0(\eta)~~,
\eea
where: 
\be
\label{g_0}
g_0(\eta)=\lim_{\rho\rightarrow0^+}{g(\rho,\eta)}=
g\sqrt{\frac{3}{2}}\frac{[\Phi-(\eta-\lambda)\Xi]
sign(\eta-\lambda)}{\Xi^2-\Phi\Theta}~~.
\ee
Up to the factor $\mp 3 g$, the 
second equation in (\ref{bd_flow0}) describes the one-dimensional 
gradient flow of the function:
\be
\label{W_0}
W_0(\eta):=\lim_{\rho\rightarrow 0^+}{W(\rho,\eta)}=
\frac{1}{\sqrt{6}}\frac{|\eta-\lambda|}{\Phi(\eta)}~~
\ee
with respect to the limiting metric:
\be
\label{metric_0}
g^{(0)}_{\eta\eta}=\lim_{\rho\rightarrow 0^+}{g_{\eta\eta}}
=\frac{\Xi^2-\Phi\Theta}{\Phi^2}~~.
\ee
This induced metric blows up on the interval $(\eta_0,\infty)$ precisely at
the points $\eta=y_j$ ($j=0\dots k$), but this is a coordinate 
singularity. Note that (\ref{metric_0}) is
continuous on $(\eta_0,\infty)- \{y_0\dots y_k\}$. It is also
clear\footnote{This can be checked directly by noticing that
$\sqrt{g^{(0)}_{\eta\eta}}$ blows up like $|\eta-y_j|^{-1/2}$ for
$\eta\rightarrow y_j$. Hence the distance
$\int_{y_j}^{y_{j-1}}{\sqrt{g^{(0)}_{\eta\eta}} d\eta}$ stays finite. It also
follows from the fact that the metric of \cite{CS} is adapted to the
toric fibration $M\rightarrow \Delta_M$, which restricts to 
the two-sphere $S_j$ over each interval
$I_j$.} that the length of each interval $(y_j,y_{j-1})$ is
finite with respect to this metric for $j=1\dots k$.  The intervals
$(\eta_0,y_k)$ and $(y_0,+\infty)$ have infinite length since they
bound the conformal infinity $Z$.

Since the region of $M$ sitting above each interval
$I_j:=(y_j,y_{j-1})$ $(j=1\dots k$) is the 2-sphere $S_j$ (a component
of the exceptional divisor), it is clear that flows of type
(\ref{bd_flow0}) lift to flows in $M_+$ which are entirely contained
inside some $S_j$.  The intersection matrix (\ref{imatrix}) shows that
the dual graph of $S_1\dots S_k$ is a chain, so $S_j$ touches only
$S_{j-1}$ and $S_{j+1}$ for $j=2\dots k-1$. Let $\chi_j$ be a
coordinate along the uncollapsed circle above $I_j$.  Since $W$ and
the metric (\ref{metric}) are independent of the fiber coordinates,
the flow equations (\ref{flow}) require $\chi_j=const$ (this can also
be seen from equations (\ref{Flow}), since $\chi_j$ are certain linear
combinations of the angular coordinates $\phi,\psi$). Hence the flow
proceeds along the sphere $S_j$ at some fixed angular value $\chi_j$
(figure 4).  Since such flows are restricted to lie in the exceptional
divisor, we shall call them {\em divisorial flows}.

As explained after relations (\ref{flow}), the sign in equations 
(\ref{bd_flow0}) must be switched when the flow passes through a noncritical 
zero of $W$. This means that the divisorial 
flow equations can be written in the form:
\bea
\label{bd_flow}
\frac{dU}{dt}&=&-\epsilon g W_c\nn\\
\frac{d\eta}{dt}&=&\epsilon g_c(\eta)~~,
\eea
where:
\be
\label{W_c}
W_c(\eta):=\frac{1}{\sqrt{6}}\frac{\eta-\lambda}{\Phi(\eta)}~~
\ee
and:
\be
\label{g_c}
g_c(\eta)=
g\sqrt{\frac{3}{2}}\frac{\Phi-(\eta-\lambda)\Xi}{\Xi^2-\Phi\Theta}~~,
\ee
where $\epsilon\in \{-1,1\}$ is now {\em constant} along each given flow 
\footnote{For the examples of Section 5, 
we shall take all divisorial flows to have $\epsilon=+1$.}.

\begin{center}
\vbox{
\begin{minipage}[t]{3in}
\begin{center}
\scalebox{0.5}{\input{chain.pstex_t}}
\end{center}
\end{minipage}

\vskip 0.2in

\begin{minipage}[t]{3in}
\begin{center}
{\bf Figure 4:} {\footnotesize Picture of divisorial flows for the case $k=4$.}
\end{center}
\end{minipage}}
\end{center}

\subsection{General properties of divisorial flows}
To understand the general properties of the flow (\ref{bd_flow}), 
we must analyze the quantities $\Phi,\Psi$ and $\Xi$. 
Notice that the first function can be written in the form \cite{CS}:
\be
\Phi(\eta)=m_j\eta-n_j~~{\rm~for~}\eta\in I_j=(y_j,y_{j-1})~~.
\ee
Using this observation, it is easy to see that $\Phi$ has exactly one
zero, namely $\eta=\eta_0=q/p$ (the point where $Z$ meets the axis
$\rho=0$, see figure 5).  Moreover, $\Phi$ is strictly greater than
zero for $\eta>\eta_0$ (the boundary of the domain of interest $D_+$)
and smaller than zero for $\eta<\eta_0$ (the boundary of the
complementary domain $D_-$). (In fact, $\Phi$ coincides with the limit
$\lim_{\rho\rightarrow 0^+}{\sqrt{\rho}F}$).  In particular, the point
$\eta_0$ is a conformal infinity for the one-dimensional metric
(\ref{metric_0}), as expected from the fact that the latter is the
restriction of (\ref{metric}) to the real axis.  Since we are
interested in the domain $D_+$, we shall restrict to $\eta>\eta_0$ in
what follows.

\begin{center}
\vbox{
\begin{minipage}[t]{3in}
\begin{center}
\mbox{\epsfxsize=2.2in \epsffile{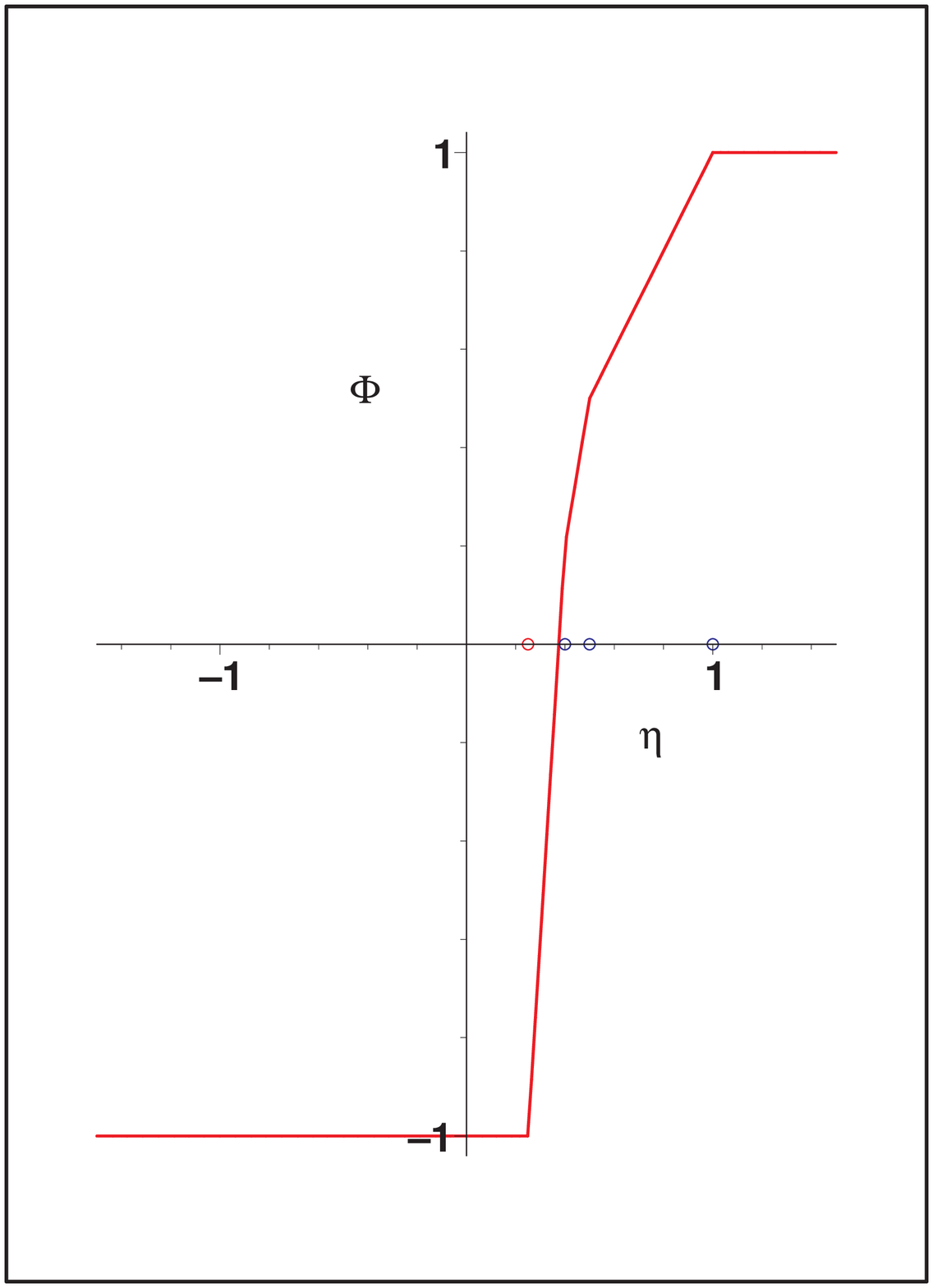}}
\end{center}
\end{minipage}

\vskip 0.2in

\begin{minipage}[t]{4in}
\begin{center}
{\bf Figure 5:} {\footnotesize The function $\Phi$ for the model 
$(p,q)=(8,3)$.}
\end{center}
\end{minipage}}
\end{center}

Expression (\ref{W_0}) shows that $W_0(\eta)$ has a zero at $\eta=\lambda$, 
which will belong to the region of interest only when $\lambda>\eta_0$. 
Notice that: 
\be
\label{W_0prime}
\frac{dW_0}{d\eta}=-|\eta-\lambda|\frac{\Xi}{\Phi^2}+
\frac{sign(\eta-\lambda)}{\Phi}~~{\rm for}~~\eta>\eta_0~~,
\ee
where we used the relation $\frac{d\Phi}{d\eta}=\Xi$. While this
quantity does not vanish at $\eta=\eta_j$, the gradient field
(\ref{g_0}) does vanish there because $\Theta$ (and thus
$g^{(0)}_{\eta\eta}$ given in (\ref{metric_0})) blows up at these
points. Once again, this is a peculiarity of our coordinate system. 
It is clear from (\ref{W_0prime}) that the derivative of $W_0$
is discontinuous at $y_j$. The same is true for the gradient
field $g_0(\eta)$. Taking $\eta\in I_j:=(y_j,y_{j-1})$, we obtain:
\be
\label{W_I}
W_0(\eta)=\frac{1}{m_j\sqrt{6}}
\frac{|\eta-\lambda|}{\eta-\lambda_j}=
\frac{sign(\eta-\lambda)}{m_j\sqrt{6}}
\left[1-\frac{\lambda-\lambda_j}{\eta-\lambda_j}\right]~~
{\rm~for~}~\eta \in I_j~~.
\ee
(remember that $\lambda_j=n_j/m_j$ is the value of $\lambda$
corresponding to the isometry which fixes $S_j$). By property (b) 
of Subsection 3.1, we have $\eta_0=1/q_k>1/q_{j-1}=\lambda_j$, 
so that $\eta-\lambda_j>0$ for $\eta>\eta_0$. Thus
$W$ is constant on $(y_j,y_{j-1})$ if $\lambda=\lambda_j$, in
agreement with the fact that the isometry defined by this value of
$\lambda$ fixes the locus $S_j$. Since $\lambda_j<y_j$, we find that
$W_0$ is nonsingular (and non-negative) along this interval.  It will
be monotonous on the entire interval unless $\lambda\in
(y_j,y_{j-1})$, in which case (assuming $\lambda>\lambda_j$) 
it decreases for $\eta\in (y_j,\lambda)$
and increases for $\eta\in(\lambda,y_{j-1})$. 
It is also clear that the sign-corrected
superpotential (\ref{W_c}) is always monotonous on $I_j$ and will be
strictly increasing if $\lambda>\lambda_j$:
\be
W_c(\eta)=\frac{1}{m_j \sqrt{6}} \left( 1-\frac{\lambda-\lambda_j}
{\eta-\lambda_j} \right)~~{\rm~for~}~\eta \in I_j~~.
\ee
With this assumption on $\lambda$, $W_c$ 
is concave (has negative second derivative) on 
$I_j-\{\lambda\}$.
Finally, note that the sign-corrected gradient field (\ref{g_c})
is continuous along $I_j$ and vanishes at the 
endpoints of this interval. Due to our choice of coordinates, the 
$\eta$-derivative of $W_c$ (computed from within the 
interval) does not vanish at the endpoints of $I_j$. However, 
$g_c$ vanishes there due to the curvature singularity of the restricted 
metric. 

Equations (\ref{bd_flow}) can be integrated by quadratures on each
interval $I_j$. In fact, it is possible to obtain the general form of
the solution of the second equation along such an interval.  To find
it, let us assume that $\eta\in I_j$. Then it is easy to check that
$\Xi(\eta)=m_j$. Combining this with $\Phi(\eta)=m_j\eta-n_j$ allows
us to compute:
\bea
\Xi^2-\Phi\Theta &=&m_j^2-m_j(\eta-\lambda_j)
\sum_{i=0}^{k+1}{\frac{w_i}{|\eta-y_i|}}\nn\\
\Phi-(\eta-\lambda)\Xi&=&\lambda m_j-n_j~~,
\eea
where we used $\lambda_j=n_j/m_j$. To simplify the first expression,
write $|\eta-y_i|=\epsilon_{ij}(\eta-y_i)$, where
$\epsilon_{ij}=sign(\eta-y_i)$ equals $+1$ for $i\geq j$ and $-1$
otherwise.  Using
$\frac{\eta-\lambda_j}{\eta-y_i}=1+\frac{y_i-\lambda_j}{\eta-y_i}$, we
obtain:
\be
\Xi^2-\Phi\Theta=-m_j\sum_{i=0}^{k+1}{\epsilon_{ij}\frac{w_i(y_i-\lambda_j)}
{\eta-y_i}}~~,
\ee
where we used $\sum_{i=0}^{k+1}{w_j\epsilon_{ij}}=\sum_{i=0}^{k+1}{
w_j sign(\eta-y_i)}=\Xi=m_j$. 

This allows us to write the sign-corrected gradient in the form:
\be \label{gc}
g_c(\eta)=-g\sqrt{\frac{3}{2}}\frac{\lambda-\lambda_j}{
\sum_{i=0}^{k+1}{\epsilon_{ij}\frac{w_i(y_i-\lambda_j)}{\eta-y_i}}}~~.
\ee

The second equation in (\ref{bd_flow}) now integrates to:
\be
\label{int}
\int{d\eta \frac{1}{g_c(\eta)}}=\epsilon t~~,
\ee
with 
\be
\int{d\eta \frac{1}{g_c(\eta)}}=-\frac{1}{g(\lambda-\lambda_j)}
\sqrt{\frac{2}{3}}
\sum_{i=0}^{k+1}{\epsilon_{ij}w_i(y_i-\lambda_j)\ln|\eta-y_i|}+ {\rm
constant}~~.
\ee
Equation (\ref{int}) gives the implicit form of the solution $\eta(t)$
on the interval $I_j$:
\be \label{etat}
\sum_{i=0}^{k+1}{\epsilon_{ij}w_i(y_i-\lambda_j)\ln|\eta-y_i|}=
-g\sqrt{\frac{3}{2}}(\lambda-\lambda_j)\epsilon t~~.
\ee
To arrive at this relation, we used the freedom of performing a flow
re-parameterization $t\rightarrow t+{\rm constant}$ in order to absorb
the additive constant of integration \footnote{A similar
re-parameterization can be used to clear the denominators in the
logarithms of (\ref{etat}), an observation which will be used
repeatedly in the examples of Section 5.}  (this re-parameterization
does not affect the behavior of the flow for $t\rightarrow \pm
\infty$).

To understand the behavior of (\ref{etat}) near the endpoints of
$I_j$, let us compute: \bea
\epsilon_{jj}w_i(y_j-\lambda_j)\ln|\eta-y_j|&=&-\frac{1}{2m_j}\ln
|\eta-y_j|\nn\\
\epsilon_{j-1,j}w_i(y_{j-1}-\lambda_j)\ln|\eta-y_{j-1}|&=&
+\frac{1}{2m_j}\ln |\eta-y_j|~~, \eea where we used the first
relations in (c) of Subsection 3.1. Since $\ln (0^+)=-\infty$, this
shows that $\eta$ tends to $y_j$ for $\epsilon
(\lambda-\lambda_j)t\rightarrow -\infty$ and to $y_{j-1}$ for
$\epsilon (\lambda-\lambda_j)t\rightarrow +\infty$.  Thus 
the flow starts at one end of $I_j$ at $t=-\infty$ and reaches the
other end at $t=+\infty$.

\subsection{Flows of Randall-Sundrum type}

Recall that a flow is of Randall-Sundrum \cite{RS} type if it connects
two `IR critical points' of $W$, i.e. two critical points for which
the matrix $\partial^\Lambda \partial_\Sigma W$  
is negative semidefinite when computed on the side of 
the flow \cite{Cortes, Behrndt_review} (remember that $W$ is always 
non-negative with our conventions). 
In our case, this definition does not strictly apply, since the metric 
blows up at the critical points in our coordinates, so that 
$\partial^\Lambda \partial_\Sigma W$ will vanish there.  Instead, we 
shall retreat to the more physical definition which requires an exponentially
decreasing warp factor near the endpoints of the flow.

In our models, a divisorial flow along $I_j$ turns out to have this type if
the restricted superpotential $W_0$ attains local maxima at the
endpoints $y_j,y_{j-1}$ of the flow \footnote{This
condition assures that the sign-corrected gradient field has the
appropriate behavior in our models. Indeed, equations (\ref{bd_flow}) 
show that the solution (\ref{etat}) asymptotes to a constant near 
the endpoints of $I_j$ (since the gradient $g_c$ vanishes there), 
so that $U$ tends to $\epsilon \times 
sign W_c(y_j)\times \infty$
and  $-\epsilon \times sign W_c(y_{j-1}) \times \infty$ at the endpoints. 
Choosing $\epsilon$ such that $\epsilon (\lambda -\lambda_j) >0$, the 
solution (\ref{etat}) flows from $y_j$ to $y_{j-1}$ and $e^{2U}$ will have 
Randall-Sundrum behavior if $sign  W_c(y_j)=-\epsilon$ and 
$sign  W_c(y_{j-1})=+\epsilon$. This means that $W_c$ must change sign 
inside $I_j$, so that $W_0$ must vanish there, in which case the endpoints 
of $I_j$ are local maxima (when viewed from within $I_j$) 
for the restriction of $W_0$ to this interval.}
(by this we mean local maxima for
the restriction of $W_0$ to $I_j$, i.e.  $y_j$ is a local maximum when
`viewed from the right' and $y_{j-1}$ is a local maximum when `viewed
from the left').  As in \cite{Behrndt_review}, this condition requires
that the flow pass through a zero of $W_0$, so that it proceeds
between a minimum and a maximum of the {\em sign-corrected}
superpotential $W_c$.  In view
of the discussion above, this happens if and only if the flow
parameter $\lambda$ belongs to the interval $I_j$. Fixing $\lambda\in
(y_k,y_0)-\{y_1\dots y_{k-1}\}$, one finds a unique interval $I_j$
$(j=1\dots k$) containing $\lambda$. The flow along this interval is
of Randall-Sundrum type, while flows along the remaining intervals
connect UV/IR fixed points (figure 6).  In particular,
this proves the existence of Randall-Sundrum flows (for a certain
range of $\lambda$) for every model in our family. We also note that
flows with $\eta>y_0$ or $\eta<y_k$ will necessarily extend to the
conformal infinity $Z$, since $W$ grows to infinity there.  Such
unbounded flows correspond to domain walls which connect a degenerate
solution (associated with the conformal infinity) with one of the
$AdS_5$ solutions defined by the critical points $y_0$ or
$y_k$.  An unbounded flow of this type will pass through a zero of $W$
if $\lambda\in (\eta_0,y_0)$ or $\lambda\in (y_k,+\infty)$
respectively.  For the remainder of this paper we shall concentrate on
divisorial flows, which proceed along one of the finite intervals
$I_j$.

\begin{center}
\vbox{
\begin{minipage}[t]{3in}
\begin{center}
\scalebox{0.5}{\input{flows.pstex_t}}
\end{center}
\end{minipage}

\vskip 0.2in

\begin{minipage}[t]{6in}
\begin{center}
{\bf Figure 6:} {\footnotesize Schematic depiction of divisorial flows
for $\lambda \in (y_2,y_1)$ (without considering the angular variables
$\chi_j$ along the two-spheres). The figure shows the case $k=3$ with
$\epsilon=+1$ and $\lambda>\lambda_j$. A point is defined to be of
UV/IR type if the associated warp factor is exponentially
increasing/decreasing in $t$ as the flow approaches that point.}
\end{center}
\end{minipage}}
\end{center}

If $\lambda\in I_j=(y_j,y_{j-1})$, then the values of the superpotential 
at the endpoints of this interval are given by (\ref{W_I}):
\be
W_0(y_j)=\frac{1}{m_j\sqrt{6}}\frac{|y_j-\lambda|}{\eta-\lambda_j}~~,~~
W_0(y_{j-1})=\frac{1}{m_j\sqrt{6}}\frac{|y_{j-1}-\lambda|}{\eta-\lambda_j}~~.
\ee
Requiring $W_0(y_j)=W_0(y_{j-1})$ with $\lambda\in I_j$ gives
$\lambda=\frac{y_j+\alpha_jy_{j-1}}{1+\alpha_j}$, where
$\alpha_j:=\frac{y_j-\lambda_j}{y_{j-1}-\lambda_j}$. Using property
(c) of Subsection 3.1, we obtain
$\alpha_j=\frac{m_j-m_{j-1}}{m_{j+1}-m_j}$, which leads to the
following expression for $\lambda$:
\be
\label{lambda_RS}
\lambda=\frac{n_{j+1}-n_{j-1}}{m_{j+1}-m_{j-1}}:=\lambda^{(j)}~~.
\ee
Thus we can always choose $\lambda$ such that $W$ has equal values at
the endpoints of a Randall-Sundrum flow. With this choice of flow
parameter, we obtain:
\be \label{W_endpts}
W_0(y_j)=W_0(y_{j-1})=\frac{1}{\sqrt{6}}\frac{e_j-2}{m_{j+1}-m_{j-1}}~~,
\ee
where we used property (c) of Subsection 3.1.  We also note that
$\lambda^{(j)}-\lambda_j=\frac{2}{m_j(m_{j+1}-m_{j-1})}>0$, so that a
Randall-Sundrum flow will satisfy:
\bea
&& \eta(t)\rightarrow y_j {\rm~~~~for~} \epsilon t\rightarrow -\infty\nn\\
&& \eta(t)\rightarrow y_{j-1} {\rm~for~} \epsilon t\rightarrow +\infty~~.
\eea
In particular, $\eta$ flows from the lower to the upper end of $I_j$ 
if one takes $\epsilon=+1$. 

At a critical point of $W$, the potential (\ref{pot}) takes the form:
\be
\label{V}
{\cal V} = - 6 W^2 \, .
\ee
This nonpositive quantity gives the cosmological constant of $AdS_5$,
which is the supergravity solution at that point in the moduli
space. With the choice of isometry given in (\ref{lambda_RS}), the
flow between $y_j$ and $y_{j-1}$ interpolates between two $AdS_5$
solutions with the same value of the cosmological constant.

\section{Examples} 

\subsection{Models with $k>1$ for low values of $p$}

Models with $k=1$ and negative $c_1(M)$ are very frequent: for each
value of $p$, one obtains such a model by setting $q=1$ (this leads to
the Pedersen metrics \cite{Pedersen}, see below).  Models with
negative $c_1$ become increasingly sparse as one increases $k$.  Let
us order all models increasingly by lexicographic order in $(p,q)$.
Then it is not hard to check that a given value of $k$ is first
realized for $p/q=f_{2k+2}/f_{2k}=(3\dots 3)$, where
$f_1=1,f_2=1,f_3=2,f_4=3,f_5=5\dots$ are the Fibonacci numbers.  In
particular,

(a) $k=2$ is first realized by $(p,q)=(8,3)$

(b) $k=3$ is first realized by $(p,q)=(21,8)$

(c) $k=4$ is first realized by $(p,q)=(55,21)$

(d) $k=5$ is first realized by $(p,q)=(144,55)$~~.

It is clear that one can realize any value of $k$ (since finite minus
continued fractions represent the rationals).  We shall illustrate the
general discussion of the previous section by giving a detailed
analysis of Randall-Sundrum 
flows for the Pedersen metrics and for the models (a) and (b).

\subsection{The Pedersen-LeBrun metrics}

In \cite{Pedersen}, H.~Pedersen constructed a one-dimensional family
of $U(2)$ invariant ESD metrics of negative scalar curvature on the
unit open four-ball $B^4$. These metrics are characterized by a
parameter\footnote{One allows $m^2$ to be positive or negative, so
that $m$ may be imaginary.}  $m^2>-1$ which fixes their behavior near
the $S^3$ boundary.  In fact, the unit 3-sphere is a conformal
infinity for these spaces, if the former is endowed with the Berger
metric:
\be
ds^2=\sigma_1^2+\sigma_2^2+I_3 \, \sigma_3^2~~,
\ee 
where $I_3=\frac{1}{m^2+1}>0$ and $\sigma_i$ are three left-invariant
one-forms on $S^3=SU(2)$ such that
$d\sigma_i=\epsilon_{ijk}\sigma_j\wedge \sigma_k$.  These metrics are
smooth and complete on $B^4$ for any $m^2>-1$; their explicit form is
given in Appendix A.

When $m^2\in (-1,0)\Leftrightarrow I_3>1$, these can be continued to
complete metrics `on the other side' of $S^3$, provided that
$|m|=\frac{p-2}{p}$ for some integer $p>2$
\cite{Hitchin}.
In this case, the Pedersen ansatz and
its continuation induce (up to a $p$-fold cover)
smooth metrics on disjoint open subsets $M_-$
and $M_+$ of the minimal resolution $M={\cal O}(-p)\rightarrow \P^1$
of $\C^2/\Z_p$, where the generator of $\Z_p$ acts through:
\be
(z_1,z_2)\rightarrow (e^{2\pi i/p}z_1, e^{2\pi i/p}z_2)~~.
\ee 
These sets $M_+$ and $M_-$ are separated by a common conformal
infinity, namely the Lens space $Y=S^3/\Z_p$, carrying the conformal
structure induced by the Berger metric.  The conformal structures on
$M_+$ and $M_-$ agree along $Y$, and define a unique conformal
structure on $M$; the latter was discovered by LeBrun in the context
of scalar-flat Kahler metrics \cite{LeBrun}.

An interesting property of the Pedersen-LeBrun models is that they
interpolate between the hyperbolic metric on $B^4$ and the Bergman
metric\footnote{We remind the reader that the hyperbolic metric on
$B^4$ is the homogeneous metric on the symmetric space $SO(4,1)/SO(4)$
(also known as $EAdS_4$), while the Bergman metric is the homogeneous
metric on $SU(2,1)/U(2)$ (also known as the metric of the universal
hypermultiplet).}  \cite{Pedersen}. The hyperbolic metric arises for
$m=0\Leftrightarrow I_3=1$, while the Bergman space is obtained in
the limit $m^2\rightarrow -1 \Leftrightarrow I_3\rightarrow
\infty$. In the second case, the conformal structure induced on
$S^3$ degenerates to a left-invariant CR structure \cite{Pedersen, Hitchin}.

As shown in \cite{Hitchin}, the Pedersen-LeBrun metrics
are the only smooth and complete ESD metrics of negative scalar
curvature which admit a $U(2)$ isometry with 3-dimensional generic
orbits. They form a special subclass in Hitchin's classification of
complete and $SU(2)$ invariant ESD metrics. Since such metrics admit an 
obvious $T^2$ symmetry, they also fit into the framework of \cite{CP}.

The Pedersen-LeBrun metrics were recently considered in \cite{BA},
though in a different parameterization which originates in the work of
\cite{P}.  Since the metrics of \cite{BA} are ESD and
$U(2)$-invariant of negative scalar curvature, Hitchin's
classification assures us that they coincide with the Pedersen-LeBrun
spaces. In appendix A, we give the explicit coordinate transformations
which reduce the models of \cite{BA} to the Pedersen form.

By using their coordinates, the authors of \cite{BA} build a flow of
Randall-Sundrum type on the subset $M_+$ of $M$ for a certain choice
of gauged isometry. In fact, the relevant isometry is a Cartan $U(1)$ factor 
inside the $SU(2)$ subgroup of the decomposition 
$U(2)=U(1)\times SU(2)$. Since this is a subgroup of 
a $T^2\subset U(2)$, 
it is natural to reconsider this problem from the point of view 
of the present paper. Below, we show how the flow of \cite{BA}
can be recovered with our techniques.

In the framework of \cite{CS}, the Pedersen metrics arise for $q=1$. 
This immediately gives 
$k=1$ and $e_1=p$. Thus the exceptional divisor consists of a 
single two-sphere of self-intersection $-p$.
The condition $p\geq 3$ implements the
constraint $c_1(M)<0$. The recursion relations (\ref{mn_recursion})
give:
\be
(m_0,n_0)=(0,-1)~~,~~(m_1,n_1)=(1,0)~~,~~(m_2,n_2)=(p,1)~~.
\ee
The toric generators of $M$ for the case $p=3$ are shown in figure 7.

\vskip 0.2in

\begin{center}
\vbox{
\begin{minipage}[t]{3in}
\begin{center}
\scalebox{0.7}{\input{31Gens.pstex_t}}
\end{center}
\end{minipage}~~~~~\begin{minipage}[t]{3in}
\begin{center}
\mbox{\epsfxsize=2.02in \epsffile{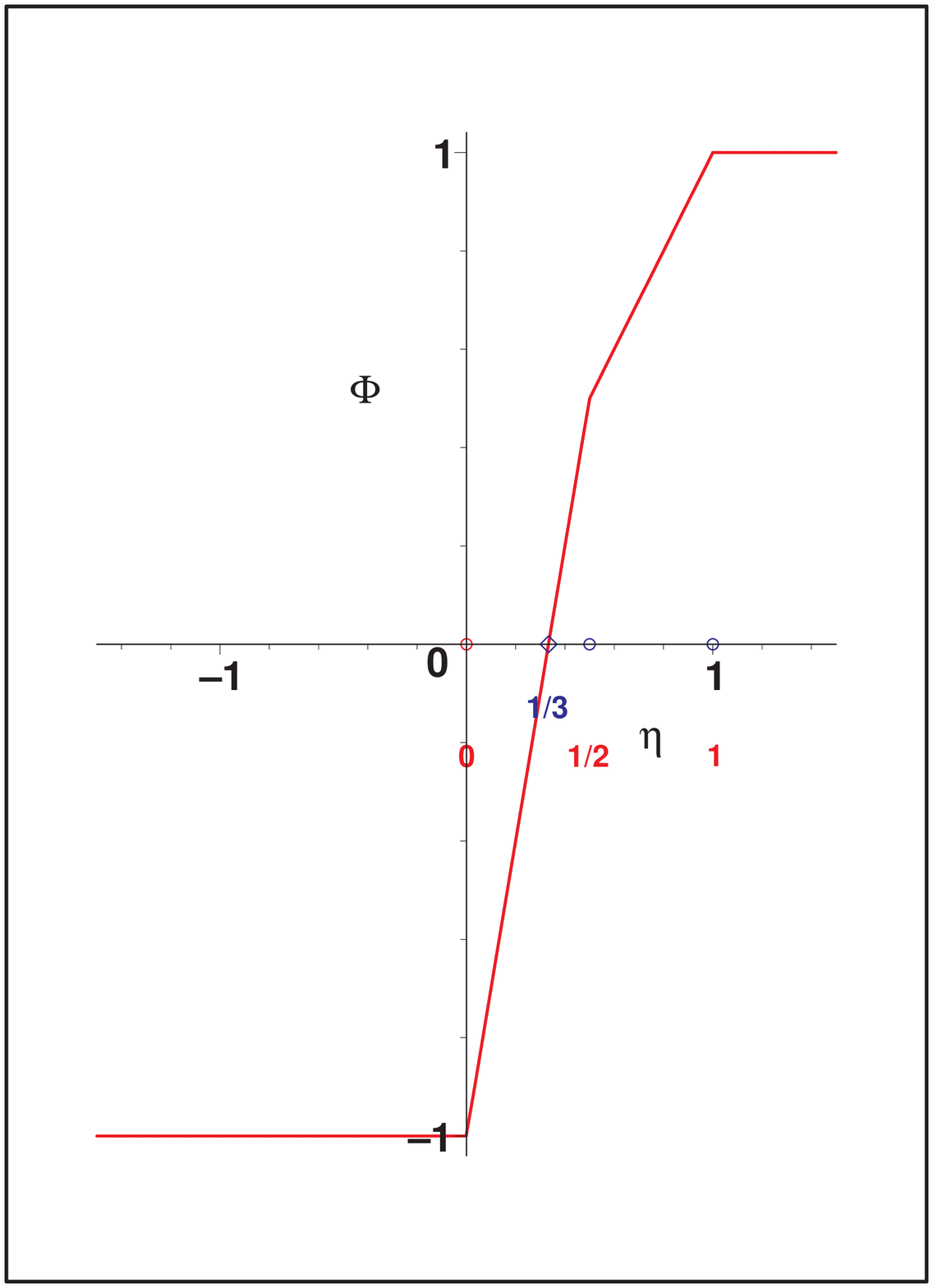}}
\end{center} 
\end{minipage}

\vskip 0.2 in

\begin{minipage}[t]{3in}\begin{center}
{\bf Figure 7:} {\footnotesize Toric generators $\nu_0,\nu_1$ and
$\nu_2$ for $(p,q)=(3,1)$. The figure does not show the vector $\nu_3=
\left[\ba{c} 0\\1 \ea\right]$ added when compactifying the minimal
resolution.}
\end{center}\end{minipage}~~~~~\begin{minipage}[t]{3in}\begin{center}
{\bf Figure 8:} {\footnotesize The function $\Phi$ for the model
$(p,q)=(3,1)$. 
The figure indicates the points $y_2=0$, $y_1=1/2$ and $y_0=1$, as well 
as the bordering value $\eta_0=1/3$. }
\end{center}\end{minipage}}
\end{center}

\vskip 0.2in 

One also has the formal pair $(m_3,n_3)=(0,1)$. This gives:
\be
y_0=1~~,~~y_1=\frac{1}{p-1}~~,~~y_2=0~~;~~ \eta_0=1/p~~
\ee
and: 
\be
w_0=-1/2~~,~~ w_1=-(p-1)/2~~,~~w_2=p/2~~.
\ee
Hence the defining function $F$ has the form:
\be
F(\rho,\eta)=\frac{1}{2}\left[-f(\rho,\eta,1)-(p-1)f(\rho,\eta,\frac{1}{p-1})+
p f(\rho,\eta,0)\right]~~.
\ee
The boundary $(1/p,\infty)$ of $D_+$
contains the points $y_0=1$ and $y_1=1/(p-1)$, while the point $y_2=0$
belongs to the boundary $(-\infty,1/p)$ of $D_-$.  The function $\Phi$
has the form:
\be
\Phi(\eta)=\frac{1}{2}\left[-|\eta-1|-(p-1)|\eta-\frac{1}{p-1}|+p|\eta|
\right]~~.
\ee
This is plotted in figure 8 (for the case $p=3$).  The isometry
defined by $\lambda^{(1)}=2/p$ has the property
$W_0(y_1)=W_0(y_0)=\frac{1}{\sqrt{6}}(1-2/p)$.  Figure 9 shows the
level lines of the superpotential for $p=3$ and
$\lambda=\lambda^{(1)}=2/3$.

\begin{center}
\vbox{
\begin{minipage}[t]{3in}
\begin{center}
\mbox{\epsfxsize=1.72in \epsffile{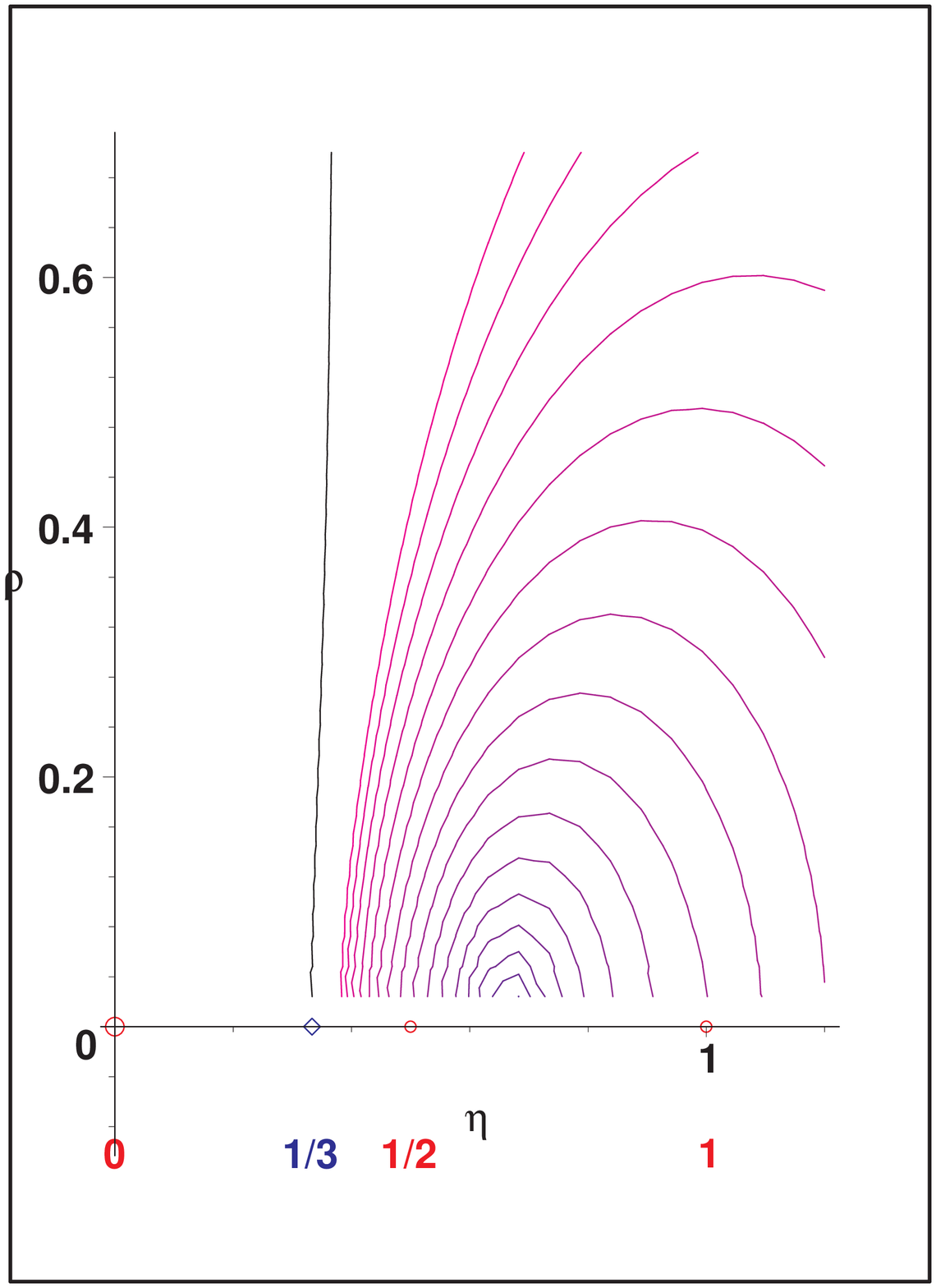}}
\end{center}
\end{minipage}~~~~~\begin{minipage}[t]{3in}
\begin{center}
\scalebox{0.6}{\begin{picture}(0,0)%
\epsfig{file=31W0.pstex}%
\end{picture}%
\setlength{\unitlength}{4144sp}%
\begingroup\makeatletter\ifx\SetFigFont\undefined%
\gdef\SetFigFont#1#2#3#4#5{%
  \reset@font\fontsize{#1}{#2pt}%
  \fontfamily{#3}\fontseries{#4}\fontshape{#5}%
  \selectfont}%
\fi\endgroup%
\begin{picture}(3759,4524)(2239,-4123)
\put(2431,-16){\makebox(0,0)[lb]{\smash{\SetFigFont{14}{16.8}{\familydefault}{\mddefault}{\updefault}
$W_0$
}}}
\end{picture}
}
\end{center}
\end{minipage}

\vskip 0.1 in

\begin{minipage}[t]{3in}
\begin{center}
{\bf Figure 9:} {\footnotesize Level lines of the superpotential for
$(p,q)=(3,1)$ and $\lambda=2/3$.  In the color version of this figure,
increasing values of $W$ are represented by a red shift in the
coloring of the level lines. The vertical bold black curve represents
the conformal infinity $Z$. In this plot, we use the convention in
which the superpotential is non-negative throughout $D_+$.}
\end{center}
\end{minipage}~~~~~\begin{minipage}[t]{3in} \begin{center}
{\bf Figure 10:} {\footnotesize The restriction of $W$ to the line 
$\rho=0$ for $\lambda=2/3$. The superpotential vanishes for $\eta=\lambda$.
Note that the directional derivatives of $W$ disagree at the points $y_1,y_0$.
The gradient of $W_0$ {\em does} vanish at these points, since the restricted 
metric blows up there. }
\end{center}
\end{minipage}}
\end{center}

\vskip 0.1in

\begin{center}
\vbox{
\begin{minipage}[t]{3in}
\begin{center}
\scalebox{0.6}{\begin{picture}(0,0)%
\epsfig{file=31metric.pstex}%
\end{picture}%
\setlength{\unitlength}{4144sp}%
\begingroup\makeatletter\ifx\SetFigFont\undefined%
\gdef\SetFigFont#1#2#3#4#5{%
  \reset@font\fontsize{#1}{#2pt}%
  \fontfamily{#3}\fontseries{#4}\fontshape{#5}%
  \selectfont}%
\fi\endgroup%
\begin{picture}(3579,4524)(2239,-4123)
\put(2746,-241){\makebox(0,0)[lb]{\smash{\SetFigFont{14}{16.8}{\familydefault}{\mddefault}{\updefault}
$g^{(0)}_{\eta\eta}$
}}}
\end{picture}
}
\end{center}
\end{minipage}~~~~~\begin{minipage}[t]{3in}
\begin{center}
\scalebox{0.6}{\begin{picture}(0,0)%
\epsfig{file=31grad.pstex}%
\end{picture}%
\setlength{\unitlength}{4144sp}%
\begingroup\makeatletter\ifx\SetFigFont\undefined%
\gdef\SetFigFont#1#2#3#4#5{%
  \reset@font\fontsize{#1}{#2pt}%
  \fontfamily{#3}\fontseries{#4}\fontshape{#5}%
  \selectfont}%
\fi\endgroup%
\begin{picture}(3714,4524)(2239,-4123)
\put(2746,-61){\makebox(0,0)[lb]{\smash{\SetFigFont{14}{16.8}{\familydefault}{\mddefault}{\updefault}
$g_0$
}}}
\end{picture}
}
\end{center}
\end{minipage}

\vskip 0.1 in

\begin{minipage}[t]{3in}
\begin{center}
{\bf Figure 11:} {\footnotesize The restricted metric
$g^{(0)}_{\eta\eta}$ for $(p,q)=(3,1)$. As explained in the text, the metric 
has a coordinate singularity at $y_1=1/2$ and $y_0=1$.}
\end{center}
\end{minipage}~~~~~\begin{minipage}[t]{3in} \begin{center}
{\bf Figure 12:} {\footnotesize The restricted gradient field
$g_0(\eta)$ for $(p,q)=(3,1)$ and $\lambda=2/3$. The vertical axis is
measured in units of the coupling constant $g$. Note the discontinuity
at $\eta=\lambda$. This is eliminated when considering the
sign-corrected field $g_c(\eta)$ of equation (\ref{gc}).} 
\end{center}
\end{minipage}}
\end{center}

The isometry fixing the interval $I_1:=(y_1,y_0)=(\frac{1}{p-1},1)$
has parameter $\lambda_1=0$. Therefore, the $\phi$-circle collapses
above this interval, while the fibration of the $\psi$-circle between
$y_1$ and $y_0$ gives the exceptional divisor $S_1$ (this is the
$\P^1$ base of the fibration ${\cal O}(-p)\rightarrow \P^1$).

On the interval $I_1$, we have $\Phi(\eta)=\eta$ and
$\Xi=m_1=1$. Therefore (see figures 10-12):
\bea
W_0(\eta)&=&\frac{1}{\sqrt{6}}\frac{|\eta-\lambda|}{\eta}~~\nn\\
g_0(\eta)&=&-\lambda g\sqrt{6}\frac{(\eta-1)\left((p-1)\eta-1\right)}{p-2}
sign(\eta-\lambda)~~.
\eea
The sign-corrected functions $W_c$ and $g_c$ are plotted in figure 13.

\begin{center}
\vbox{
\begin{minipage}[t]{3in}
\begin{center}
\scalebox{0.6}{\input{31Wc.pstex_t}}
\end{center}
\end{minipage}~~~~~\begin{minipage}[t]{3in}
\begin{center}
\scalebox{0.61}{\begin{picture}(0,0)%
\epsfig{file=31sol.pstex}%
\end{picture}%
\setlength{\unitlength}{4144sp}%
\begingroup\makeatletter\ifx\SetFigFont\undefined%
\gdef\SetFigFont#1#2#3#4#5{%
  \reset@font\fontsize{#1}{#2pt}%
  \fontfamily{#3}\fontseries{#4}\fontshape{#5}%
  \selectfont}%
\fi\endgroup%
\begin{picture}(3624,4524)(2239,-4123)
\put(4186,-331){\makebox(0,0)[lb]{\smash{\SetFigFont{17}{20.4}{\familydefault}{\mddefault}{\updefault}
$\eta$
}}}
\end{picture}
}
\end{center}
\end{minipage}

\vskip 0.2in

\begin{minipage}[t]{3in}
\begin{center}
{\bf Figure 13:} {\footnotesize Sign-corrected superpotential $W_c$ and gradient 
field $g_c$ on the interval (1/2,1) (for $\lambda=2/3$). Note continuity of 
$g_c$ and the zero of $W_c$ at $\eta=\lambda$.}
\end{center}
\end{minipage}~~~~~\begin{minipage}[t]{3in} \begin{center}
{\bf Figure 14:} {\footnotesize The flow $\eta(t)$ along the 
interval $(1/2,1)$ for $(p,q)=(3,1)$ and $\lambda=2/3$. This flow
is of Randall-Sundrum type.}
\end{center}
\end{minipage}}
\end{center}

\vskip 0.2in

The solution (\ref{etat}) takes the following form on the interval $I_1$
(if one chooses $\epsilon=+1$):
\be
\frac{1}{2}\ln|\eta-1|-\frac{1}{2}\ln |\eta-\frac{1}{p-1}|=-g\sqrt{\frac{3}{2}}
\lambda t~~.
\ee
Upon clearing denominators in the logarithms and absorbing the resulting 
constant by a shift of $t$ (see the footnote after equation (\ref{etat})),
we obtain $\ln ((p-1)\eta-1)-\ln (1-\eta)=g\lambda \sqrt{6}t$, which gives:
\bea
\label{p1flow}
\eta(t)&=&\frac{e^{-\lambda\sqrt{6}gt}+1}{(p-1)e^{-\lambda\sqrt{6}gt}+1}~~\nn\\
U(t)&=&-\frac{1}{6}\left[(p-2)\ln(e^{-\lambda\sqrt{6}gt}+1)
-(\lambda-1)\sqrt{6}gt\right]~~.
\eea
(this solution satisfies $\eta(-\infty)=y_1=\frac{1}{p-1}$ and
$\eta(+\infty)=y_0=1$). $\eta(t)$ is plotted in figure 14 for $p=3$
and $\lambda=2/3$. For $\lambda=\lambda^{(1)}=2/p$, the solution for
$U$ becomes:
\be
\label{warp}
U(t)=-\frac{p-2}{6}\ln{\left[2\cosh(\frac{\sqrt{6}}{p}gt)\right]}~~.
\ee
For each $\lambda\in I_1$, the flow (\ref{p1flow}) has Randall-Sundrum 
type and proceeds along a meridian of the two-sphere $S_1$ (figure 15).

\begin{center}
\vbox{
\begin{minipage}[t]{3in}
\begin{center}
\scalebox{0.4}{\input{Sflows.pstex_t}}
\end{center}
\end{minipage}

\vskip 0.2in

\begin{minipage}[t]{6in}
\begin{center}
{\bf Figure 15:} {\footnotesize Divisorial flows for the Pedersen
models. The flows proceed along the meridians of the $\P^1$ base of
$M$, between the two critical points located at the poles $p_1$,$p_0$.
The interval $I_1=(\frac{1}{p-1},1)$ sits along the axis connecting
these poles.}
\end{center}
\end{minipage}}
\end{center}

Let us discuss the relation with \cite{BA}. Since we don't
know the explicit coordinate transformation between the
Calderbank-Pedersen metric (\ref{metric}) and the parameterization
used in \cite{BA}, we cannot immediately compare our solution for $\eta$
with the solution for their coordinate $x$.  However, we can compare
the warp factors, since in both cases they depend only on the fifth
space-time coordinate (denoted by $\rho$ in \cite{BA} and by $t$ in
the present paper).  The solution given in equation (94) of \cite{BA}
has the form:
\be 
\label{BAwarp}
e^{2 U (\rho)} = [\cosh (2 g \rho)]^{-\frac{2}{3} (n-r_-) (n+r_+)} \, ,
\ee
where $r_{\pm} = -n \pm \sqrt{4n^2 -1}$\footnote{Equation (94) of
\cite{BA} actually reads $e^{2 U (\rho)} = [\cosh (2 g
\rho)]^{-\frac{1}{3} (n-r_-) (n+r_+)}$. The missing factor of $2$ in
the exponent is due to a typo in \cite{BA}, as one can check that
equations (91), (92) and (93) of that paper give $e^{U (\rho)} =
[\cosh (2 g\rho)]^{-\frac{1}{3} (n-r_-) (n+r_+)}$.  We also note a
missing factor of $g$ in the RHS of the first equation (92) of
\cite{BA}.}.  So $(n-r_-) (n+r_+) = (2n+\sqrt{4n^2-1}) \,
\sqrt{4n^2-1}$. Our solution (\ref{warp}) gives:
\be
\label{e2U}
e^{2 U(t^{\prime})} = 2^{\frac{2-p}{3}} 
[\cosh (2 g t^{\prime})]^{-\frac{p-2}{3}} \, ,
\ee
where we have rescaled $t^{\prime} = \frac{\sqrt{6} t}{2 p}$, which
should be identified with the coordinate $\rho$ of \cite{BA}. This
change of scale is due to the different normalization chosen in
\cite{BA} for the Killing vector (\ref{K}). Together with the power of
two prefactor in (\ref{e2U}), this can be absorbed into a constant
rescaling of the spacetime metric (\ref{ansatz}), as explained in the
observation following equation (\ref{K}).

Hence to compare the solutions we must compare the exponents in
(\ref{BAwarp}) and (\ref{e2U}). Using $p = \frac{2}{1-|m|}$ and the
relation $m^2 = -(1 - \frac{1}{4n^2})$, we obtain:
\be
p = \frac{2}{1-\sqrt{1 - \frac{1}{4n^2}}} = \frac{4n}{2n - \sqrt{4n^2-1}} \, .
\ee
Hence
\be 
p-2 = \frac{2 \, \sqrt{4n^2-1}}{2n - \sqrt{4n^2-1}} = \frac{2 \,
\sqrt{4n^2 -1} (2n + \sqrt{4n^2 - 1})}{4n^2 - (4n^2 - 1)} = 2 \,
\sqrt{4n^2 -1} (2n + \sqrt{4n^2 - 1}) \, .  
\ee
This agrees with the exponent of (\ref{BAwarp}).

\

{\bf Observation} It may seem that our restricted superpotential
disagrees with that plotted in Figure 2 of \cite{BA}. However, the
authors of \cite{BA} use a parameterization in which the $\P^1$ base
of $M$ (=the exceptional divisor $S_1$) is mapped onto the complex
plane after removing the point at infinity (this plane is
parameterized by $x+iy$, where $x,y$ are the real
coordinates on $\P^1$ used in Section 5 of \cite{BA}). In this case,
the critical points (corresponding to the poles of $\P^1$) sit at
$y=0$ and $x=\pm 1$, and the flow of \cite{BA} proceeds along the real
axis between $x_-$ and $x_+$. In our description, the segment $[x_-,
x_+]$ corresponds to a meridian connecting the poles, while the entire
real axis corresponds to the full circle $S^1\subset \P^1=S^2$ which
contains this meridian. When following this circle, one covers our
interval $[y_1,y_0]$ {\em twice}: once as one passes from $x_-$ to
$x_+$, and once again when going from $x_+$ to the point at infinity
and back to $x_-$ from the other side of the real axis.  The first
step gives the restriction of $W_c$ to $[y_0,y_1]$, which should map to the
restriction of the potential of \cite{BA} to the interval $[x_-,x_+]$
after an appropriate change of coordinates.
A moment's thought shows that the second step should be responsible for the
rest of figure 2 of \cite{BA}: 
when following the other meridian determined by our big circle, one 
obtains the restriction of figure 2 of \cite{BA} to 
$(-\infty,x_-]\cup [x_+,\infty)$ upon applying the appropriate change of 
coordinates to a second copy of the restriction of $W_c$ to our 
interval $[y_1,y_0]$.

\subsection{The model (p,q)=(8,3)}

In this case, the Calderbank-Singer metric is defined on the minimal
resolution of the orbifold $\C^2/\Z_8$ with action:
\be
(z_1,z_2)\rightarrow (e^{\pi i/4} z_1, e^{3\pi i/4}z_2)~~.
\ee
The intersection matrix of the exceptional $\P^1$'s is given by 
(\ref{imatrix}), with the integers $e_j$ determined by the continued fraction
expansion:
\be
p/q=8/3=(3,3)=3-\frac{1}{3}~~.
\ee
The minimal resolution has negative $c_1$ since $e_1,e_2\geq 3$. The model 
has $k=2$, which gives four distinguished points $y_0\dots y_{k+1}$ 
on the conformal boundary of ${\cal H}^2$. Solving the recursion relations 
(\ref{mn_recursion}) gives: $(m_0,n_0)=(0,-1)$, $(m_1,n_1)=(1,0)$, 
$(m_2,n_2)=(3,1)$ and $(m_3,n_3)=(8,3)$. As explained above, one 
also adds the formal values $(m_4,n_4)=(0,1)$. The quantities $y_0\dots y_3$ 
and $w_0\dots w_3$ are given by (\ref{yw_def}):
\bea
(y_0\dots y_3)&=&(1,1/2,2/5,1/4)~~\nn\\
(w_0\dots w_3)&=&(-1/2, -1, -5/2,4)~~.
\eea
The conformal boundary $Z=\{F=0\}$ meets the real axis in the point 
$\eta_0=q/p=3/8$. The spheres $S_1,S_2$ are fixed by the isometries defined by 
the following values of $\lambda$:
\be
(\lambda_0\dots \lambda_4)=(-\infty, 0, 1/3, 3/8,\infty)~~.
\ee
The toric generators $\nu_0\dots \nu_3$ are shown in figure \ref{83gens} of 
Section 3. The exceptional divisor of such models consists of two spheres
$S_1$ and $S_2$, associated with the intervals $I_1=(y_1,y_0)=(1/2,1)$ 
and $I_2=(y_2,y_1)=(2/5,1/2)$ (figure 16). 
Figures 17 and 18 show 
the level lines of the superpotential for 
$\lambda=2/3$ and $\lambda=3/7$. 

\begin{center}
\vbox{
\begin{minipage}[t]{3in}
\begin{center}
\scalebox{0.4}{\input{83div.pstex_t}}
\end{center}
\end{minipage}

\vskip 0.2in

\begin{minipage}[t]{5in}
\begin{center}
{\bf Figure 16:} {\footnotesize The exceptional divisor for $(p,q)=(8,3)$.
The Randall-Sundrum flows constructed below proceed along the meridians
of one of the spheres.}
\end{center}
\end{minipage}}
\end{center}

\begin{center}
\vbox{
\begin{minipage}[t]{3in}
\begin{center}
\mbox{\epsfxsize=2.2in \epsffile{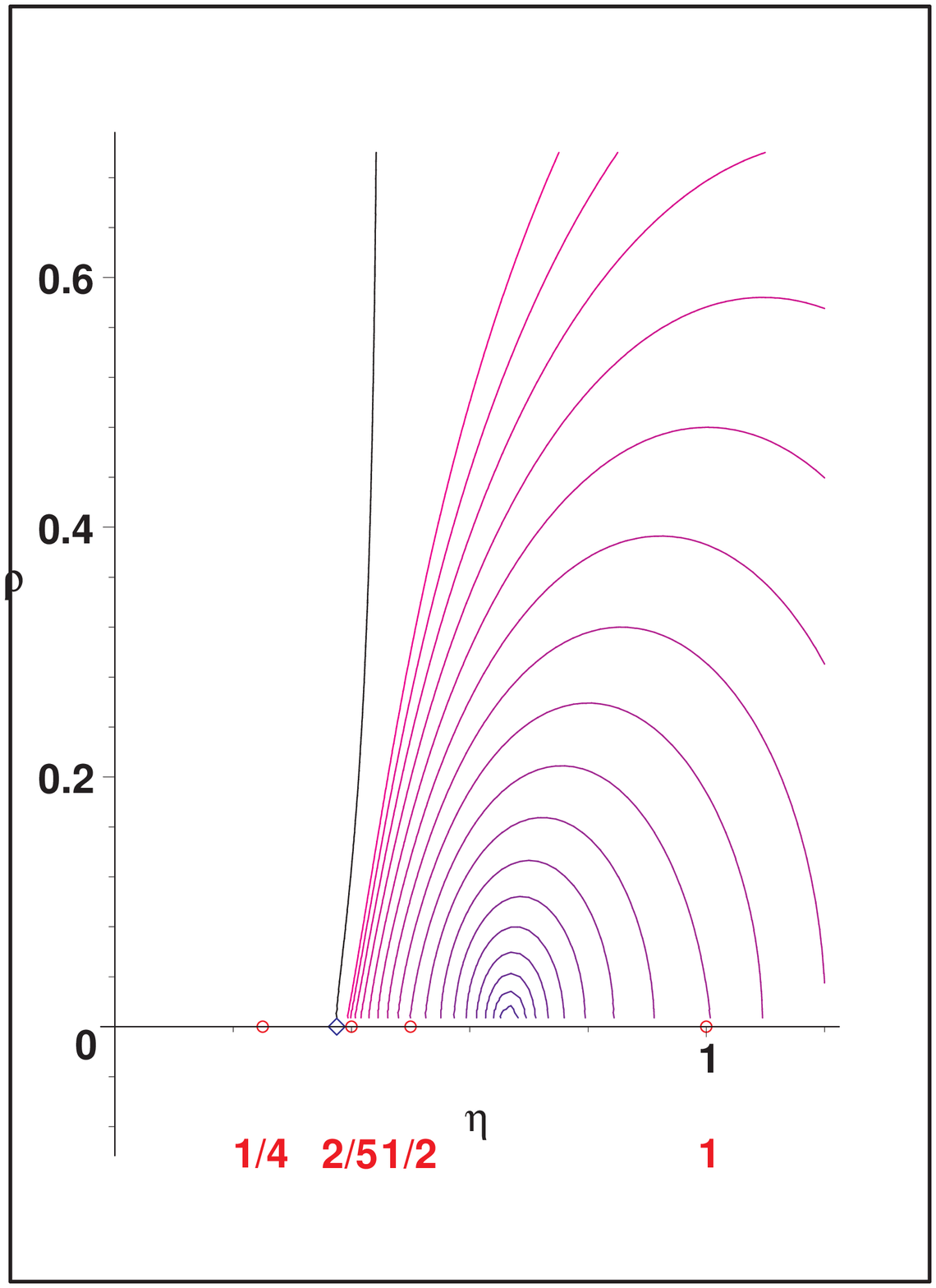}}
\end{center}
\end{minipage}~~~~~\begin{minipage}[t]{3in}
\begin{center}
\mbox{\epsfxsize=2.2in \epsffile{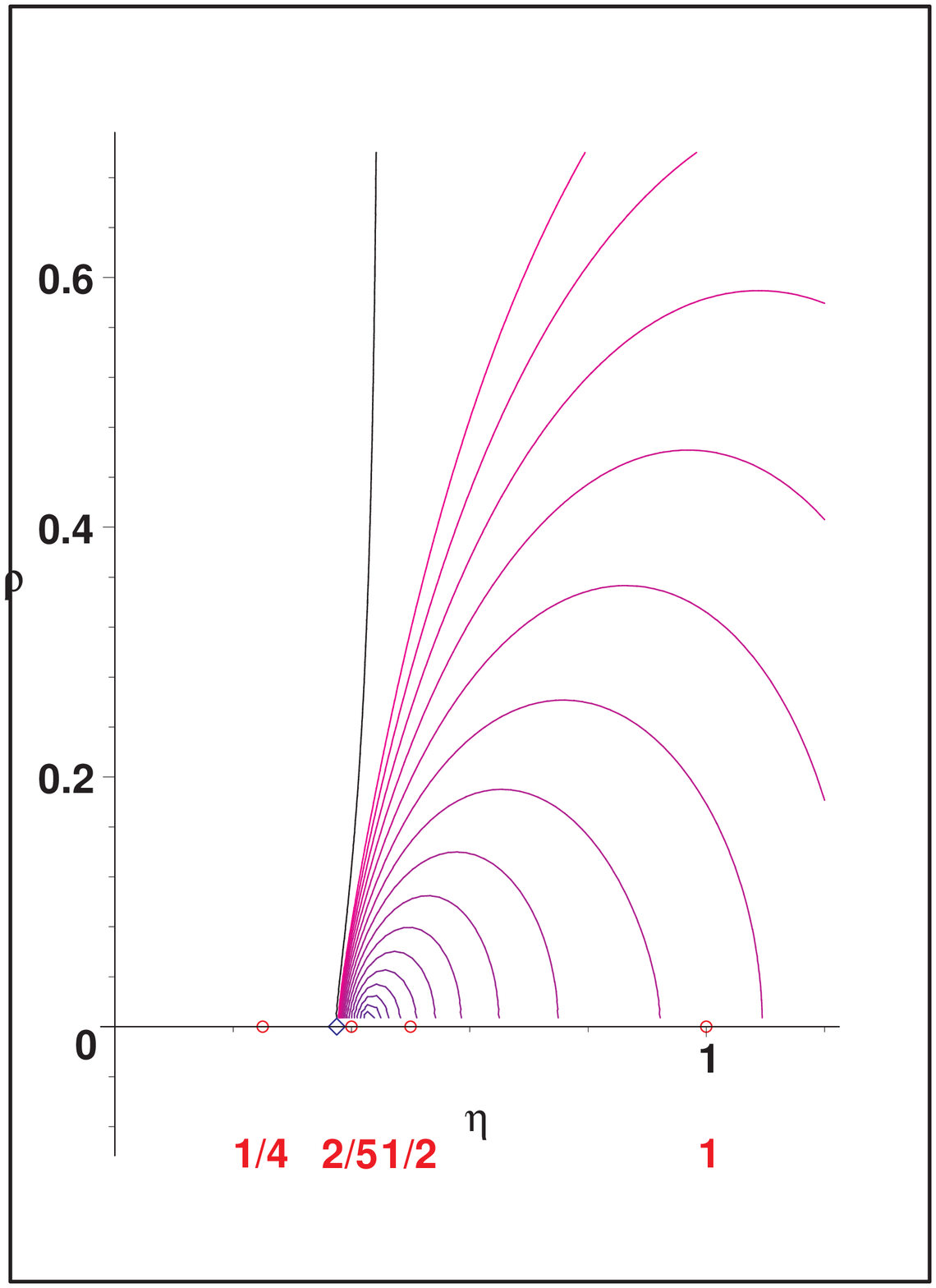}}
\end{center}
\end{minipage}

\vskip 0.2in

\begin{minipage}[t]{3in}
\begin{center}
{\bf Figure 17:} {\footnotesize Level lines of the superpotential
 for $(p,q)=(8,3)$ and $\lambda=2/3$. 
Increasing values of $W$ are represented by a red shift in 
the coloring of the level lines. The vertical bold black curve represents 
the conformal infinity $Z$. The superpotential is taken to be 
non-negative throughout $D_+$.}
\end{center}
\end{minipage}~~~~~\begin{minipage}[t]{3in} \begin{center}
{\bf Figure 18:} {\footnotesize Level lines of the superpotential for 
$(p,q)=(8,3)$ and $\lambda=3/7$.}

\end{center}
\end{minipage}}
\end{center}

\vskip 0.2in

\begin{center}
\vbox{
\begin{minipage}[t]{3in}
\begin{center}
\scalebox{0.575}{\input{83Wc1.pstex_t}}
\end{center}
\end{minipage}~~~~~\begin{minipage}[t]{3in}
\begin{center}
\scalebox{0.6}{\begin{picture}(0,0)%
\epsfig{file=83sol1.pstex}%
\end{picture}%
\setlength{\unitlength}{4144sp}%
\begingroup\makeatletter\ifx\SetFigFont\undefined%
\gdef\SetFigFont#1#2#3#4#5{%
  \reset@font\fontsize{#1}{#2pt}%
  \fontfamily{#3}\fontseries{#4}\fontshape{#5}%
  \selectfont}%
\fi\endgroup%
\begin{picture}(3489,4524)(2239,-4123)
\put(3646,-196){\makebox(0,0)[lb]{\smash{\SetFigFont{14}{16.8}{\familydefault}{\mddefault}{\updefault}
$\eta$
}}}
\end{picture}
}
\end{center}
\end{minipage}

\vskip 0.2in

\begin{minipage}[t]{3in}
\begin{center}
{\bf Figure 19:} {\footnotesize Sign-corrected superpotential $W_c$ 
and gradient field $g_c$ on the interval (1/2,1) for $\lambda=2/3$.}
\end{center}
\end{minipage}~~~~~\begin{minipage}[t]{3in} \begin{center}
{\bf Figure 20:} {\footnotesize Flow along the interval 
$I_1=(1/2,1)$ for $\lambda=2/3$.}
\end{center}
\end{minipage}}
\end{center}

To find a Randall-Sundrum flow on the interval $I_1=(1/2,1)$, we pick 
a value for $\lambda\in I_1$ and compute the form of 
the gradient field $g_0(\eta)$ of equation (\ref{g_0}) along this interval.
Taking into account the sign prefactors, one obtains:
\be
g_c(\eta)=
-\lambda g \sqrt{6}\frac{(\eta-1)(2\eta-1)(5\eta-2)(4\eta-1)}
{8\eta^2+5\eta-4}~~.
\ee
The solution of (\ref{bd_flow}) is given implicitly by the equation:
\be
\ln(1-\eta)-\ln(2\eta-1)-2\ln(5\eta-2)+2\ln(4\eta-1)= -\lambda\sqrt{6}gt~~,
\ee
where we used the freedom to shift $t$ in order to absorb a term 
$\ln\left(\frac{25}{8}\right)$ from the left hand side. 
This solution is plotted in figure 20 
for the choice $\lambda=2/3$, 
which ensures $-W_c (1/2)=W_c (1)=\frac{1}{3\sqrt{6}}$.  The full flow 
proceeds along some fixed meridian of the sphere $S_1$.

For $\lambda,\eta\in I_2=(2/5, 1/2)$, one obtains:
\be
g_c(\eta)=
-g \sqrt{6}\frac{(3\lambda-1)(\eta-1)(2\eta-1)(5\eta-2)(4\eta-1)}
{64\eta^2-59\eta+13}~~,
\ee
with sign-corrected superpotential $W_c(\eta)=\frac{1}{\sqrt{6}}
\frac{\eta-\lambda}{3\eta-1}$.

The divisorial flow $\eta(t)$ is given by the equation:
\be
-2\ln(1-\eta)-\ln(1-2\eta)+\ln(5\eta-2)+2\ln(4\eta-1)=
(3\lambda-1)\sqrt{6} gt~~.
\ee
This flow is plotted in figure 22, for $\lambda=3/7$. 
With this value of $\lambda$, one has $-W_c(2/5)=W_c(1/2)=\frac{\sqrt{6}}{42}$.
The full flow proceeds along some fixed meridian of the sphere $S_2$.

\begin{center}
\vbox{
\begin{minipage}[t]{3in}
\begin{center}
\scalebox{0.545}{\input{83Wc2.pstex_t}}
\end{center}
\end{minipage}~~~~~\begin{minipage}[t]{3in}
\begin{center}
\scalebox{0.6}{\begin{picture}(0,0)%
\epsfig{file=83sol2.pstex}%
\end{picture}%
\setlength{\unitlength}{4144sp}%
\begingroup\makeatletter\ifx\SetFigFont\undefined%
\gdef\SetFigFont#1#2#3#4#5{%
  \reset@font\fontsize{#1}{#2pt}%
  \fontfamily{#3}\fontseries{#4}\fontshape{#5}%
  \selectfont}%
\fi\endgroup%
\begin{picture}(3624,4524)(2239,-4123)
\put(3061,-511){\makebox(0,0)[lb]{\smash{\SetFigFont{14}{16.8}{\familydefault}{\mddefault}{\updefault}
$\eta$
}}}
\end{picture}
}
\end{center}
\end{minipage}

\vskip 0.2in

\begin{minipage}[t]{3in}
\begin{center}
{\bf Figure 21:} {\footnotesize Sign-corrected superpotential and gradient 
field on the interval $I_2=(2/5,1/2)$ (for $\lambda=3/7$).}
\end{center}
\end{minipage}~~~~~\begin{minipage}[t]{3in} \begin{center}
{\bf Figure 22:} {\footnotesize Flow along the interval $I_2=(2/5,1/2)$ for 
$\lambda=3/7$.}
\end{center}
\end{minipage}}
\end{center}

\vskip 0.2in

\subsection{The model (p,q)=(21,8)}

In this case, we have $k=3$ and $e_1=e_2=e_3=3$. The recursions 
(\ref{mn_recursion}) give:
\bea
(m_0\dots m_5)&=&(0, 1, 3, 8, 21, 0)\nn\\
(n_0\dots n_5)&=&(-1, 0, 1, 3, 8, 1)~~.
\eea
The curve $Z$ intersects the line $\rho=0$ at $\eta_0=8/21$. 
One easily computes:
\bea
(y_0\dots y_4)&=&(1,1/2,2/5,5/13,1/3)~~\nn\\
(w_0\dots w_4)&=&(-1/2, -1, -5/2, -13/2, 21/2)~~\nn\\
(\lambda_1\dots \lambda_4)&=&(0, 1/3,3/8,8/21)~~.
\eea
The function $\Phi$ for this model is shown in figure 24 below.
The points $(y_0\dots y_3)=(1, 1/2, 2/5, 5/13)$ belong to $D_+$, 
while $y_4=1/3$ 
(added when compactifying the minimal resolution)
belongs to $D_-$. The boundary of $D_+$ contains three finite length intervals 
$I_1 = (1/2,1)$, $I_2 = (2/5,1/2)$ and $I_3 = (5/13,2/5)$. One obtains flows 
of Randall-Sundrum type along these intervals (with equal values of $W$ at the 
endpoints of the flow) for the following values of $\lambda$ 
(see equation (\ref{lambda_RS})):
\be
\label{lambda_RS21_8}
(\lambda^{(1)},\lambda^{(2)},\lambda^{(3)})=(2/3,3/7,7/18)~~.
\ee
Equation (\ref{etat}) gives the following expressions for 
the Randall-Sundrum flows along these intervals:
\bea 
\label{sol21_8}
I_1 \, : \,\,\,\,\, \ln (1-\eta )&-&\ln (2\eta -1) -2\ln(5\eta
-2)- 5\ln (13\eta -5) + 7\ln (3\eta -1) \nn \\  
= - g\sqrt{6}\lambda t \nn \\ 
I_2\, : \, 2\ln (1-\eta )&+&\ln (1-2\eta) -\ln(5\eta -2)- 2\ln (13\eta
-5) \nn \\ &=& -
g\sqrt{6}(3\lambda -1) t \nn \\ I_3 \, : \, 5\ln (1-\eta )&+&2\ln
(1-2\eta) -\ln(2-5\eta)- \ln (13\eta -5) - 7\ln (3\eta -1) \nn \\ &+&
= - g\sqrt{6}(8\lambda -3) t \, .
\eea
The geometry of these flows is shown in figure 23.
Figures 25, 26 and 27 plot the solutions (\ref{sol21_8}) for the values of $\lambda$ 
given in (\ref{lambda_RS21_8}). From (\ref{W_endpts}) we find the 
superpotentials: 
$W_0^{(I_1)} (1) = \frac{1}{3 \sqrt{6}} \, , \, 
W_0^{(I_2)} (\frac{1}{2}) = \frac{1}{7 \sqrt{6}} \, , \, 
W_0^{(I_3)} (\frac{2}{5}) = \frac{1}{18 \sqrt{6}}$.

\begin{center}
\vbox{
\begin{minipage}[t]{3in}
\begin{center}
\scalebox{0.4}{\input{21_8div.pstex_t}}
\end{center}
\end{minipage}

\vskip 0.2in

\begin{minipage}[t]{4in}
\begin{center}
{\bf Figure 23:} {\footnotesize Randall-Sundrum flows for $(p,q)=(21,8)$.}
\end{center}
\end{minipage}}
\end{center}

\begin{center}
\vbox{
\begin{minipage}[t]{3in}
\begin{center}
\mbox{\epsfxsize=1.8in \epsffile{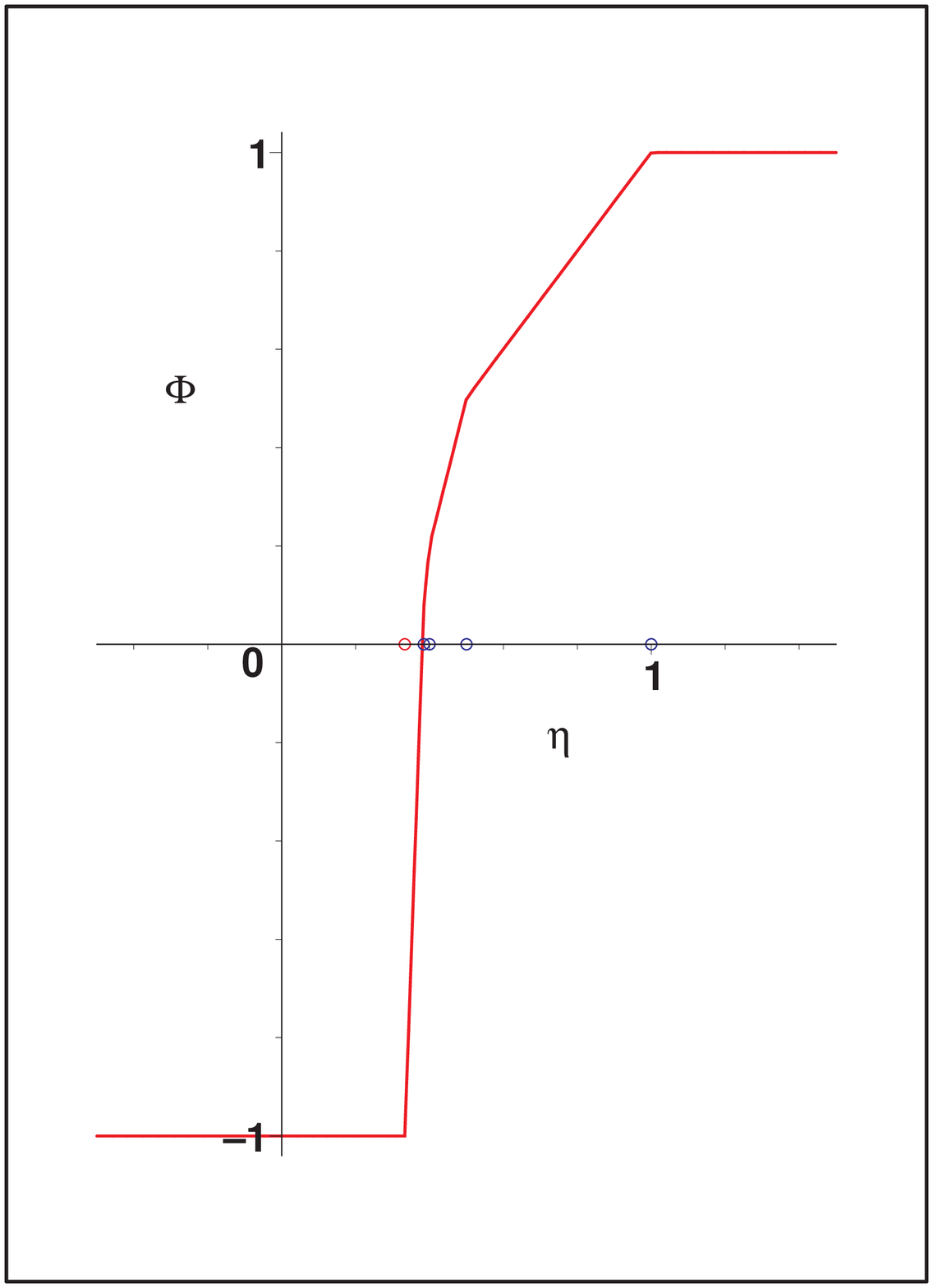}}
\end{center}
\end{minipage}~~~~~\begin{minipage}[t]{3in}
\begin{center}
\scalebox{0.6}{\begin{picture}(0,0)%
\epsfig{file=21_8sol1.pstex}%
\end{picture}%
\setlength{\unitlength}{4144sp}%
\begingroup\makeatletter\ifx\SetFigFont\undefined%
\gdef\SetFigFont#1#2#3#4#5{%
  \reset@font\fontsize{#1}{#2pt}%
  \fontfamily{#3}\fontseries{#4}\fontshape{#5}%
  \selectfont}%
\fi\endgroup%
\begin{picture}(3579,4569)(2239,-4123)
\put(2656,-196){\makebox(0,0)[lb]{\smash{\SetFigFont{14}{16.8}{\familydefault}{\mddefault}{\updefault}
$\eta$
}}}
\end{picture}
}
\end{center}
\end{minipage}

\vskip 0.2in

\begin{minipage}[t]{3in}
\begin{center}
{\bf Figure 24:} {\footnotesize The function $\Phi$ for $(p,q)=(21,8)$.}
\end{center}
\end{minipage}~~~~~\begin{minipage}[t]{3in} \begin{center}
{\bf Figure 25:} {\footnotesize  Flow of Randall-Sundrum type for $\lambda=2/3$.}
\end{center}
\end{minipage}}
\end{center}

\begin{center}
\vbox{
\begin{minipage}[t]{3in}
\begin{center}
\scalebox{0.6}{\begin{picture}(0,0)%
\epsfig{file=21_8sol2.pstex}%
\end{picture}%
\setlength{\unitlength}{4144sp}%
\begingroup\makeatletter\ifx\SetFigFont\undefined%
\gdef\SetFigFont#1#2#3#4#5{%
  \reset@font\fontsize{#1}{#2pt}%
  \fontfamily{#3}\fontseries{#4}\fontshape{#5}%
  \selectfont}%
\fi\endgroup%
\begin{picture}(3624,4479)(2239,-4123)
\put(3106,-196){\makebox(0,0)[lb]{\smash{\SetFigFont{14}{16.8}{\familydefault}{\mddefault}{\updefault}
$\eta$
}}}
\end{picture}
}
\end{center}
\end{minipage}~~~~~\begin{minipage}[t]{3in}
\begin{center}
\scalebox{0.6}{\begin{picture}(0,0)%
\epsfig{file=21_8sol3.pstex}%
\end{picture}%
\setlength{\unitlength}{4144sp}%
\begingroup\makeatletter\ifx\SetFigFont\undefined%
\gdef\SetFigFont#1#2#3#4#5{%
  \reset@font\fontsize{#1}{#2pt}%
  \fontfamily{#3}\fontseries{#4}\fontshape{#5}%
  \selectfont}%
\fi\endgroup%
\begin{picture}(3669,4569)(2239,-4123)
\put(5356,164){\makebox(0,0)[lb]{\smash{\SetFigFont{14}{16.8}{\familydefault}{\mddefault}{\updefault}
$\eta$
}}}
\end{picture}
}
\end{center}
\end{minipage}

\vskip 0.2in

\begin{minipage}[t]{3in}
\begin{center}
{\bf Figure 26:} {\footnotesize Flow of Randall-Sundrum type for $\lambda=3/7$.}
\end{center}
\end{minipage}~~~~~\begin{minipage}[t]{3in} \begin{center}
{\bf Figure 27:} {\footnotesize  Flow of Randall-Sundrum type for $\lambda=7/18$.}
\end{center}
\end{minipage}}
\end{center}

\section{Conclusions}

We constructed an infinite family of gauged N=2
supergravities in 5 dimensions admitting Randall-Sundrum flows. The
matter content of these models consists of a single hypermultiplet
described by the complete and smooth quaternion-Kahler four-manifolds
constructed recently by Calderbank and Singer.  These metrics are
defined on an open subset of the minimal resolution of a cyclic
singularity, and have negative scalar curvature provided that this
resolution has negative definite first Chern class.  They admit a
$T^2$'s worth of isometries and generalize the well-known Pedersen
metrics \cite{Pedersen}, which admit a larger $U(2)$ symmetry.

By using the coordinates of \cite{CP}, we found a very simple
expression for the superpotential obtained by gauging one of the toric
isometries, and showed that the associated flow preserves each of the
two-spheres which form the irreducible components of the exceptional
divisor. Moreover, the restriction of the flow to the exceptional
divisor can be described by a simple ordinary differential equation,
which can be integrated by quadratures. We also showed that the
restriction of the flow to each given sphere is of Randall-Sundrum
type for an appropriate range of choices of the gauged isometry. The
models obtained in this manner allow for an arbitrary number
of critical points.

For a few models in this class, we performed a detailed study of such
`divisorial' flows and gave explicit constructions in the 
Randall-Sundrum case. Finally, we pointed out that the models
considered recently in \cite{BA} are the well-known Pedersen metrics,
albeit discussed in a different parameterization. They form the simplest
case of the family analyzed in the present paper. Clearly our models
give an infinite number of counterexamples to the no-go theorems
mentioned in the introduction. As in \cite{BA}, the reason is very
simple: such theorems were formulated for supergravity coupled
only to vector/tensor 
multiplets or they assumed homogeneity of the hypermultiplet
moduli space. Finally, the result of \cite{MN} relies 
on non-positivity assumptions for the scalar potential  
which fail to hold along our flows: one has ${\cal V}>0$ when the
flow passes through the zero of $W$. 

It would be interesting to study possible RG flow interpretations of
the domain wall solutions interpolating between UV and IR critical
points, which were also found in the present paper. 
A more challenging question is whether our models
embed consistently in string/M-theory.

\acknowledgments{The authors thank 
M. Ro\v{c}ek for support and interesting conversations. This
work was supported by the Research Foundation under NSF grant
PHY-0098527.}

\appendix

\section{Relation between different coordinate systems for the Pedersen 
metrics}

The Pedersen metric \cite{Pedersen} has the form:

\be \label{Ped} g = \frac{1}{(1-\rho^2)^2} \left[ \frac{1+m^2
\rho^2}{1+m^2 \rho^4} d\rho^2 + \rho^2 (1+m^2 \rho^2) (\sigma_1^2 +
\sigma_2^2) + \frac{\rho^2 (1+m^2 \rho^4)}{1+m^2 \rho^2} \sigma_3^2
\right] \, , \ee 
where $\sigma_i$ are the standard $SU(2)$
left-invariant one-forms, obeying $d\sigma_i = \epsilon_{ijk} \,
\sigma_j \wedge \sigma_k$: 
\bea 
\sigma_1 &=& \frac{1}{2} (- \sin \psi
\, d \theta + \cos \psi \, \sin \theta \, d\phi) \nn \\ \sigma_2 &=&
\frac{1}{2} (\cos \psi \, d\theta + \sin \psi \, \sin \theta \, d\phi)
\nn \\ \sigma_3 &=& \frac{1}{2} (d\psi + \cos \theta \, d\phi) 
\eea
with 
$0\le \theta <\pi$, \, $0 \le \phi < 2\pi$ and $0 \le \psi < 4\pi$. Clearly
$\sigma_1^2 + \sigma_2^2 = \frac{1}{4} (d\theta^2 + \sin^2 \theta \,
d\phi^2)$.

Behrndt and Dall'Agata give several different parameterizations of
their metric and also the coordinate transformations between them. For
convenience, let us take the parameterization (43) of
\cite{BA}\footnote{This form of the metric is valid only when the 
parameter  $\kappa$ of \cite{BA} equals one, 
which is exactly the case we are interested in.}: \bea
\label{bd} ds^2 &=& \frac{dr^2}{V(r)} + V(r) (d\tau + 2n \cos \theta
\, d\phi)^2 + (r^2-n^2) (d\theta^2 + \sin^2 \theta \, d\phi^2) \, ,
\nn \\ V(r) &=& \frac{r-n}{r+n} \, [(r+n)^2 + 1 - 4n^2] \, .  \eea 
One obtains (\ref{Ped}) from (\ref{bd}) by the following coordinate
transformation: 
\be 
\label{ctr} r = \frac{\rho^2}{2n (1-\rho^2)} + n
\, , \qquad \tau = 2n \psi \, 
\ee 
and the identification $m^2 =\frac{1}{4n^2} - 1$. 
Actually doing this results in four times the metric
(\ref{Ped}). But this was to be expected since the scalar curvature of
(\ref{bd}) is normalized to $-12$, whereas Pedersen \cite{Pedersen}
normalizes the cosmological constant to $-12$ (i.e. the scalar
curvature to $-12 \times 4 = -48$). The extra factor of 4 coming as a
result of the coordinate transformation (\ref{ctr}) accounts for that
difference.

Taking \, $n = 1/2$ \, in the metric (\ref{bd}) and performing the
coordinate transformation (\ref{ctr}) again with \, $n = 1/2$ \, one
obtains: \be \label{eads4} \frac{4}{(1-\rho^2)^2} \left[ d \rho^2 +
\rho^2 (\sigma_1^2 + \sigma_2^2 + \sigma_3^2) \right] \, , \ee which
is the metric of Euclidean $AdS_4$ space, normalized so that the
scalar curvature is $-12$. The same metric (\ref{eads4}) (without the
factor of $4$) one obtains from (\ref{Ped}) upon setting $m = 0$.

Another form of the Pedersen metric is \cite{Hitchin}: \be
\label{Hitch} g = \frac{1}{(\cos \rho^{\, \prime} - M \sin \rho^{\,
\prime})^2} \left[ (1 + M \cot \rho^{\, \prime}) [d\rho^{\, \prime \,
2} + 4 \sin^2 \rho^{\, \prime} \,\, (\sigma_1^2 + \sigma_2^2)] +
\frac{4M^2}{1+M \cot \rho^{\, \prime}} \, \sigma_3^2 \right].  \ee Its
relation to (\ref{Ped}) is given by: \be \label{HPtr} \cos \rho^{\,
\prime} = \frac{1}{\sqrt{1+\frac{\rho^4}{M^2}}} \, , \qquad M =
\frac{1}{m} \, .  \ee Again one obtains four times (\ref{Ped}) due to
different normalizations between \cite{Pedersen} and \cite{Hitchin}.


\begin{thebibliography}{150}
\bibitem{LOSW}{A.~Lukas, B.~Ovrut, K.~Stelle, D.~Waldram, {\em The
Universe as a Domain Wall}, Phys. Rev. {\bf D59} (1999) 086001,
hep-th/9803235} 
\bibitem{HW}{P.~Horava, E.~Witten, {\em Heterotic and
Type I String Dynamics from Eleven Dimensions}, Nucl. Phys. {\bf B460}
506, hep-th/9510209} 
\bibitem{Witten_sh}{E.~Witten, 
{\em Strong coupling  expansion of Calabi-Yau compactification},
Nucl.Phys. {\bf B471}(1996) 135, hep-th/9602070.} 
\bibitem{Mald}{J.~Maldacena, {\em The Large N
Limit of Superconformal Field Theories and Supergravity},
Adv. Theor. Math. Phys. {\bf 2} (1998) 231, hep-th/9711200}
\bibitem{RS}{L.~Randall, R.~Sundrum, {\em An alternative to
compactification}, Phys. Rev. Lett. {\bf 83} (1999) 4690,
hep-th/9906064.}  \bibitem{KL}{R.~Kallosh, A.~Linde, {\em
Supersymmetry and the Brane World}, JHEP {\bf 02}, (2000) 005,
hep-th/0001071} 
\bibitem{CvB}{K.~Behrndt, M.~Cveti\v{c}, {\em Anti-de Sitter vacua of gauged supergravities with 8 supercharges}, Phys. Rev. {\bf D61} (2000) 101901, hep-th/0001159}
\bibitem{GibL}{G.~Gibbons, N.~Lambert, {\em Domain
Walls and Solitons in Odd Dimensions}, Phys. Lett. {\bf 488} (2000)
90, hep-th/0003197} \bibitem{MN}{J.~Maldacena, C.~Nunez, {\em
Supergravity Description of Field Theories on Curved Manifolds and a
no-go Theorem}, Int. J. Mod. Phys. {\bf A16} (2001) 822,
hep-th/0007018} \bibitem{BA}{K.~Behrndt, G.~Dall'Agata, {\em Vacua of
N=2 gauged supergravity derived from non-homogeneous quaternionic
spaces}, Nucl.Phys. B627 (2002) 357-380, hep-th/0112136.}
\bibitem{CP}{D.~M.~J. Calderbank, H.~Pedersen, {\em Selfdual Einstein
metrics with torus symmetry}, math-DG/0105263.}
\bibitem{dWRV}{B.~de~Wit, M.~Ro\v{c}ek, S.~Vandoren, {\em Hypermultiplets, Hyperkahler Cones and Quaternion-Kahler Geometry}, JHEP 0102 (2001) 039, hep-th/0101161}
\bibitem{CD}{A.~Ceresole, G.~Dall'Agata {\em General matter coupled
$N=2$, $D=5$ gauged supergravity}, hep-th/0004111.}
\bibitem{CDKP}{A.~Ceresole, G.~Dall'Agata, R.~Kallosh, A.~Van~Proeyen
{\em Hypermultiplets, Domain Walls and Supersymmetric Attractors},
hep-th/0104056.} 
\bibitem{Cortes}{D.~V.~Alekseevsky, V.~Cortes, C.~Devchand, A.~Van Proeyen, 
{\em Flows on quaternionic-Kaehler and very special real manifolds}, 
hep-th/0109094.}
\bibitem{DF}{R.~D'Auria, S.~Ferrara {\em On
Fermion Masses, Gradient Flows and Potential in Supersymmetric
Theories}, hep-th/0103153.} 
\bibitem{Hitchin}{N.~Hitchin, {\em Twistor spaces, Einstein metrics and 
isomonodromic deformations}, J. Diff. Geom. {\bf 42}(1995)30-112.}
\bibitem{GL}{K. Galicki, H. B. Lawson, Jr, {\em Quaternionic reduction and 
quaternionic orbifolds}, Math. Ann. {\bf 282} (1988) 1-21.}
\bibitem{BS}{R. L. Bryant, S. M. Salamon, {\em On the construction of some 
complete metrics with exceptional holonomy}, Duke Math. J. {\bf 58}
(1989)3, 829.}
\bibitem{toric}{L.~Anguelova, C.~I.~Lazaroiu, 
{\em M-theory compactifications on certain `toric' cones of $G_2$ holonomy}, 
hep-th/0204249.} 
\bibitem{metrics}{L. Anguelova, C.I. Lazaroiu, 
{\em M-theory on `toric' $G_2$ cones and its type II reduction}, 
hep-th/0205070.}
\bibitem{Acharya_Witten}{B.~Acharya, E.~Witten, {\em Chiral Fermions
from Manifolds of $G_2$ Holonomy}, hep-th/0109152.} 
\bibitem{GPP}{G.~W.~Gibbons, D.~N.~Page, C.~N.~Pope, 
{\em Einstein metrics on $S^3$, 
$\R^3$ and $\R^4$ bundles}, Commun. Math. Phys {\bf 127} (1990), 529-553.}
\bibitem{CS}{D.~Calderbank, M.~A.~Singer, {\em Einstein metrics and 
complex singularities}, math-DG/0206229.} 
\bibitem{EH}{T.~Eguchi, A.~Hanson, {\em Asymptotically Flat Self-Dual
Solutions to Euclidean Gravity}, Phys. Lett. {\bf 74B} (1978) 249;
G.~Gibbons, S.~Hawking, 
{\em Gravitational Multi-Instantons}, Phys. Lett. {\bf 78B} (1978) 430.}
\bibitem{Pedersen}{H.~Pedersen, {\em Einstein metrics, spinning top motions 
and monopoles}, Math. Ann. {\bf 274} (1986) 35-39.}
\bibitem{LeBrun}{C.~R.~LeBrun, {\em Counterexamples to the generalized positive action conjecture}, Commun. Math. Phys, {\bf 118} (1988) 591-596.}
\bibitem{P}{D.~N.~Page, C.~N.~Pope, {\em Inhomogeneous Einstein metrics on 
complex line bundles}, Class. Quant. Grav. {\bf 4} (1987) 213.}
\bibitem{BPV}{W.~Barth, C.~Peters, A.V. de Wen, 
{\em Compact Complex Surfaces}, Springer-Verlag, 1984.}
\bibitem{Danilov}{V.~Danilov, {\em The geometry of toric varieties}, 
Russian Math. Surveys {\bf 33}:2(1978), 97.}
\bibitem{Oda}{T. Oda, {\em Convex bodies and algebraic geometry.  
An introduction to the theory of toric varieties}. 
Ergebnisse der Mathematik und ihrer Grenzgebiete, 15. 
Springer-Verlag, Berlin, 1988.}
\bibitem{Fulton}{W.~Fulton, {\em Introduction to toric varieties}, 
Annals of mathematics studies  no. {\bf 131}, Princeton University Press, 
1993.}
\bibitem{Audin}{M.~Audin, {\em The topology of torus actions on symplectic
manifolds}, Progress in Math. {\bf 93}, Birkhauser, 1991.}
\bibitem{Behrndt_review}{K. Behrndt, {\em 
Domain walls and flow equations in supergravity}, Fortsch.Phys. {\bf 49} 
(2001) 327-338,  hep-th/0101212.}
\bibitem{GRW}{M.~Gunaydin, L.~J.~Romans, N.~P.~Warner, {\em Gauged
$N=8$ Supergravity in Five Dimensions}, Phys. Lett. {\bf B154} (1985)
268.}
\bibitem{PPN}{M.~Pernici, K.~Pilch, P.~van~Nieuwenhuizen, {\em Gauged
$N=8$ $D=5$ Supergravity}, Nucl. Phys. {\bf B259} (1985) 460.}
\bibitem{KPW}{A.~Khavaev, K.~Pilch, N.~P.~Warner, {\em New vacua of
gauged $N=8$ supergravity in five dimensions}, Phys. Lett. {\bf B487}
(2000) 14, hep-th/9812035.}
\bibitem{Ca}{M.~Cvetic et al., {\em Embedding AdS Black Holes in ten
and Eleven Dimensions}, Nucl. Phys. {\bf B558} (1999) 96,
hep-th/9903214.}
\bibitem{WN}{B.~de~Wit, H.~Nicolai, {\em The Consistency of the $S^7$
Truncation in $d=11$ Supergravity}, Nucl. Phys. {\bf B281} (1987)
211.}
\bibitem{NVN}{H.~Nastase, D.~Vaman, P.~van~Nieuwenhuizen, {\em
Consistent nonlinear KK reduction of 11d supergravity on $AdS_7\times
S^4$ and self-duality in odd dimensions}, Phys. Lett. {\bf B469}
(1999) 96, hep-th/9905075; {\em Consistency of the $AdS_7\times S^4$
reduction and the origin of self-duality in odd dimensions},
Nucl. Phys. {\bf B581} (2000) 179, hep-th/9911238.}
\bibitem{FGPW}{D.~Z.~Freedman, S.~S.~Gubser, K.~Pilch, N.~P.~Warner,
{\em Renormalization Group Flows from Holography -- Supersymmetry and
a c-Theorem}, Adv. Theor. Math. Phys. {\bf 3} (1999) 363,
hep-th/9904017.}
\bibitem{GPPZ}{L.~Girardello, M.~Petrini, M.~Porrati, A.~Zaffaroni,
{\em The Supergravity Dual of $N=1$ Super Yang-Mills Theory}, 
Nucl. Phys.{\bf B569} (2000) 451, hep-th/9909047.}
\end{thebibliography}
\end{document}